\documentclass[nofootinbib,twocolumn,showpacs,preprintnumbers,superscriptaddress,amsmath,amssymb,10pt]{revtex4-1}

\usepackage{natbib}

\usepackage[T1]{fontenc}

\usepackage{ulem}

\usepackage[pdftex]{graphicx}% Include figure files
\usepackage{bm}% bold math

\usepackage{mathcomp}
\usepackage{amsmath}

\usepackage{fancyhdr}
\usepackage{textcomp}

\usepackage[justification=raggedright,singlelinecheck=false]{caption}

\begin{document}

\normalem

\title{Light-matter Interactions in Twodimensional Transition Metal Dichalcogenides: 
Dominant Excitonic Transitions in mono- and few-layer MoX$_2$ and Band Nesting}
\author{Roland Gillen}\email{r.gillen@tu-berlin.de}
\affiliation{Institut f\"ur Festk\"orperphysik, Technische Universit\"at Berlin, Hardenbergstr. 36, 10623 Berlin, Germany}
\author{Janina Maultzsch}
\affiliation{Institut f\"ur Festk\"orperphysik, Technische Universit\"at Berlin, Hardenbergstr. 36, 10623 Berlin, Germany}

\date{\today}

\begin{abstract}
We report \textit{ab initio} calculations of the dielectric function of six mono- and bilayer molybdenum dichalcogenides based in a Bethe Salpeter equation+G$_0$W$_0$ (BSE@G$_0$W$_0$)ansatz, focussing on the excitonic transitions dominating the absorption spectrum up to an excitation energy of 3.2\,eV. Our calculations suggest that switching chalcogen atoms and the strength of interlayer interactions should affect the detailed composition of the high 'C' peaks in experimental optical spectra of molybdenum dichalcogenides and cause a significant spin-orbit-splitting of the contributing excitonic transitions in monolayer MoSe$_2$ and MoTe$_2$. This can be explained through changes in the electronic dispersion around the Fermi energy along the chalcogen series S$\rightarrow$Se$\rightarrow$Te that move the van-Hove singularities in the density of states of the two-dimensional materials along the \textit{$\Gamma$}-\textit{K} line in the Brillouin zone. Further, we confirm the distinct interlayer character of the '\textsl{C}' peak transition in few-layer MoS$_2$ that was predicted before from experimental data and show that a similar behaviour can be expected for MoSe$_2$ and MoTe$_2$ as well.
\end{abstract}

\maketitle

\section{Introduction}
The experimental realization of Graphene in 2004\,\cite{graphene-2004,novoselov-2005} opened the gates to a whole scientific field of quasi-twodimensional materials with the promise of novel applications and peculiar physical phenomena by virtue of their reduced dimensionality. Among these, the transition metal dichalcogenides (TMD) of molybdenum and tungsten are an interesting addition due to their intrinsic semiconducting nature and high chemical stability both in bulk and few-layer phases. Analogously to graphite, these materials assume a hexagonal crystal structure with stacked layers of quasi-twodimensional atomic layers that are bound together through van-der-Waals forces. The strong spin-orbit interaction and missing inversion symmetry in the monolayer materials cause a significant split of the valence band edge with an associated valley pseudospin and magnetic moment. This gives rise to coupled spin- and valley physics in these materials\,\cite{spin-valley} that manifest, for instance, in optically induced spin and valley polarizations due to valley-dependent circular dichroism and a combination of the spin and valley Hall effects.%
{\let\thefootnote\relax\footnote{\\{\copyright}2016 IEEE. Personal use of this material is permitted. Permission from IEEE must be obtained for all other uses, in any current or future media, including reprinting/republishing this material for advertising or promotional purposes, creating new collective works, for resale or redistribution to servers or lists, or reuse of any copyrighted component of this work in other works.}}

At the same time, being indirect semiconductors in their bulk forms, the attenuation of interlayer interaction causes a transition from indirect to direct band gap materials between bi- and monolayer TMD structures and a strong enhancement of photoluminescence quantum yield for decreasing material thickness\,\cite{mak-2010,splendiani-2010,MoSe2WSe2MoS2-PL,WS2-PL,MoS2-tunable-PL,scheuschner-2014}. These findings suggest a strong light-matter interaction in these 
low-dimensional materials, which, apart from the insights from the scientific point-of-view, suggest possible novel applications of two-dimensional TMDs in thin and flexible optoelectronic devices, such as photodiodes\,\cite{led1,led2} and photodetectors\,\cite{photodetector1}, single-photon emitters\,\cite{single-photon3,single-photon1,%single-photon4,
single-photon2} or in spin- and valleytronic devices\,\cite{zeng-2012}. However, their quasi-2D nature is a double-edged sword in the sense that the low thickness limits the absorption efficiency and optical quantum yields of the TMD materials. A number of strategies have thus recently been reported to increase the light-matter interaction and thus the absorption and emission efficiency by placing TMD materials in optical microcavities and enhancing the coupling of light to excitonic\,\cite{schwarz-2014,liu-2015-light-matter,garstein-2015} or trionic\,\cite{baeten-2015} dipoles through formation of cavity polaritons. Another approach is tuning the light-matter interaction through resonant exciton-plasmon coupling in hybrid systems of TMDs in plasmonic lattices, \emph{e.g.} gold nanoantennas\,\cite{kern-2015} or silver nanodisk arrays\,\cite{butun-2015,liu-2016-exciton-plasmon}. 

At the heart of these interesting physics are the excited states in the form of excitons and trions. The low-dimensionality of the TMD systems allows for long-range Coulomb interaction channels through the adjacent environment of reduced electric field screening and gives rise to tightly bound exciton and trion states that dominate experimental optical spectra of doped\,\cite{MoS2-trions} and undoped\,\cite{mak-2010} few-layer TMDs and have binding energies that are an order of magnitude higher than in conventional bulk systems. On the other hand, the corresponding excitonic wavefunctions are fairly extended, with radii on the order of several nm, and can be well described in the framework of Mott-Wannier theory\,\cite{lambrecht-2012}. This mixed Wannier-Mott and Frenkel character of excitons is well-known for carbon nanotubes\,\cite{nt-excitons} and other two-dimensional materials\,\cite{Cudazzo-esxciton-bandstructure}.
However, the low thickness renders the materials highly sensitive to the environment. Another problem arises from peak broadening from the strong electron-phonon coupling\,\cite{li-elphon} and makes determination of the exciton binding energy as difference between exciton peak energy and the band edge difficult. Correspondingly, the derived binding energies for the prominent sub-gap excitons in molybdenum and tungsten TMDs depend highly on the details of the experimental setup and, to the best of our knowledge, experimental studies of the exciton binding energy have been so far limited to monolayer TMDs. A variety of strategies for the measurement of the excitonic binding energies have been reported, including \emph{(i)} application of a modified, nonhydrogenic, Rydberg model to one-photon photoluminescence excitation (PLE) spectra\,\cite{MoS2-exciton-binding-3, WS2-exciton-binding-1}, and determination of the electronic band edge through \emph{(ii)} scanning tunneling spectroscopy (STS)\,\cite{MoS2-exciton-binding-1,ugeda-MoSe2-excitons}, \emph{(iii)} photocurrent spectroscopy\,\cite{MoS2-exciton-binding-2} and \emph{(iv)} two-photon PLE\,\cite{WSe2-exciton-binding1} spectroscopy.
For monolayer MoS$_2$, this yielded experimental binding energies of the lowest-energy exciton between 220 and 570\,meV\,\cite{MoS2-exciton-binding-1,MoS2-exciton-binding-2,MoS2-exciton-binding-3}, while the reported binding energies for other monolayer Mo and W dichalcogenides  are typically in the range of 0.3-1.0\,eV\,\cite{ugeda-MoSe2-excitons,yang-MoTe2-exciton-binding,WS2-exciton-binding-1,plechinger-WS2,WS2-WSe2exciton-binding, WSe2-exciton-binding1}. Beyond the exciton binding energy, several groups have 
reported experimental studies on the peculiar exciton and trion dynamics and the photocarrier relaxation pathways\,\cite{kozawa-TMD-photocarriers-bandnesting,vega-mayoral-2016,Malic-WS2,trion-dynamics}, that provide insights into the interplay of excited quasiparticles with the underlying electronic band structure. This is beneficial for understanding, \emph{e.g.}, the mechanism of charge separation of interlayer excitons in photodevices based on heterostructures of stacked TMD materials\,\cite{hong-hetero}.

On the theoretical side, a variety of studies of the optical properties of quasi-2D TMDs have been reported based on \textit{ab initio} calculations employing the excitonic Bethe-Salpeter equation and various analytical approaches. Here, a main focus was the interpretation of experimental optical spectra and the theoretical derivation of relevant exciton binding energies, Bohr radii, and quasi-particle band gaps. In a similar fashion to experiments, the reported values show a significant dependence on the computational details: the obtained binding energies of the lowest-energy exciton in MoS$_2$ and similar systems are in the range of 0.2-1.0\,eV\,\cite{qiu-2013,qiu-erratum,wirtz-MoS2-excitons,Ramasubramian-excitons,huser-MoS2,shi-2012,gunnar-MoS2,komsa-excitons,palumno-2015,kidd-2016}, with the absolute peak positions moving accordingly. One cause of this significant scattering is the peculiar anisotropic screening of the Coulomb interaction in two-dimensional systems that obtains a distinct dependence on the in-plane momentum transfer \textit{q} in the vicinity of the \textit{$\Gamma$} point and requires sufficiently dense \textit{q}-point grids or analytical modeling to yield reliable results\,\cite{huser-MoS2}. Based on this, recent \textit{ab initio} studies investigated the exciton radiative lifetimes in MoS$_2$/WS$_2$ and MoSe$_2$/WSe$_2$ heterobilayer structures\,\cite{palumno-2015} and the temperature-dependent effect of electron-phonon coupling on the simulated optical spectra and obtained good quantitative agreement with experiments. On the other hand, analytic methods based on tight-binding or parametrized Hamiltonians lack the flexibility of first principles approaches, but possess the necessary computational simplicity to go beyond static descriptions of isolated excitons and to study higher charge carrier complexes, such as trions\,\cite{Berkelbach-trions, komsa-2015, kidd-2016} and biexcitons\,\cite{komsa-2015, kidd-2016}, exciton dynamics\,\cite{Malic-WS2}, carrier density dependent optical spectra\,\cite{steinhoff-MoS2} or the excitonic bandstructure\,\cite{MoS2-exciton-bandstructure}. 

However, while the nature of the strongest excitonic transitions has been established for monolayer MoS$_2$, possible effects of the changed electronic dispersions compared to monolayer MoS$_2$ from different chalcogen species (as in MoSe$_2$ and MoTe$_2$) or interlayer interactions have so far not been addressed, to the best of our knowledge. Insights into the spatial distribution of the exciton wavefunctions in monolayer and few-layer materials and the differences compared to the well-studied monolayer MoS$_2$ could contribute to the understanding of interlayer excitons in TMD heterostructures and few-layer systems and related experimental results, such as recently discovered interlayer resonant Raman modes in few-layer MoS$_2$\,\cite{scheuschner-interlayer-modes}. Based on theoretical simulations, we thus analyse in this paper the contribution of excitonic states to the absorption spectra in mono- and bilayer MoS$_2$, MoSe$_2$, and MoTe$_2$ and their distributions in real and reciprocal space. We show that the changes in electronic dispersion due to interlayer interactions and different chalcogen atoms move the dominant region of band nesting between valence and conduction band and affect the composition of the dominant transitions close to the electronic band gap. Further, we confirm the existence of interlayer excitons in all three bilayer materials due to the out-of-plane character of the conduction band orbitals at the band nesting points. %The computational details can be found in the supplementary information.

\begin{figure*}
\centering
\includegraphics*[width=\textwidth]{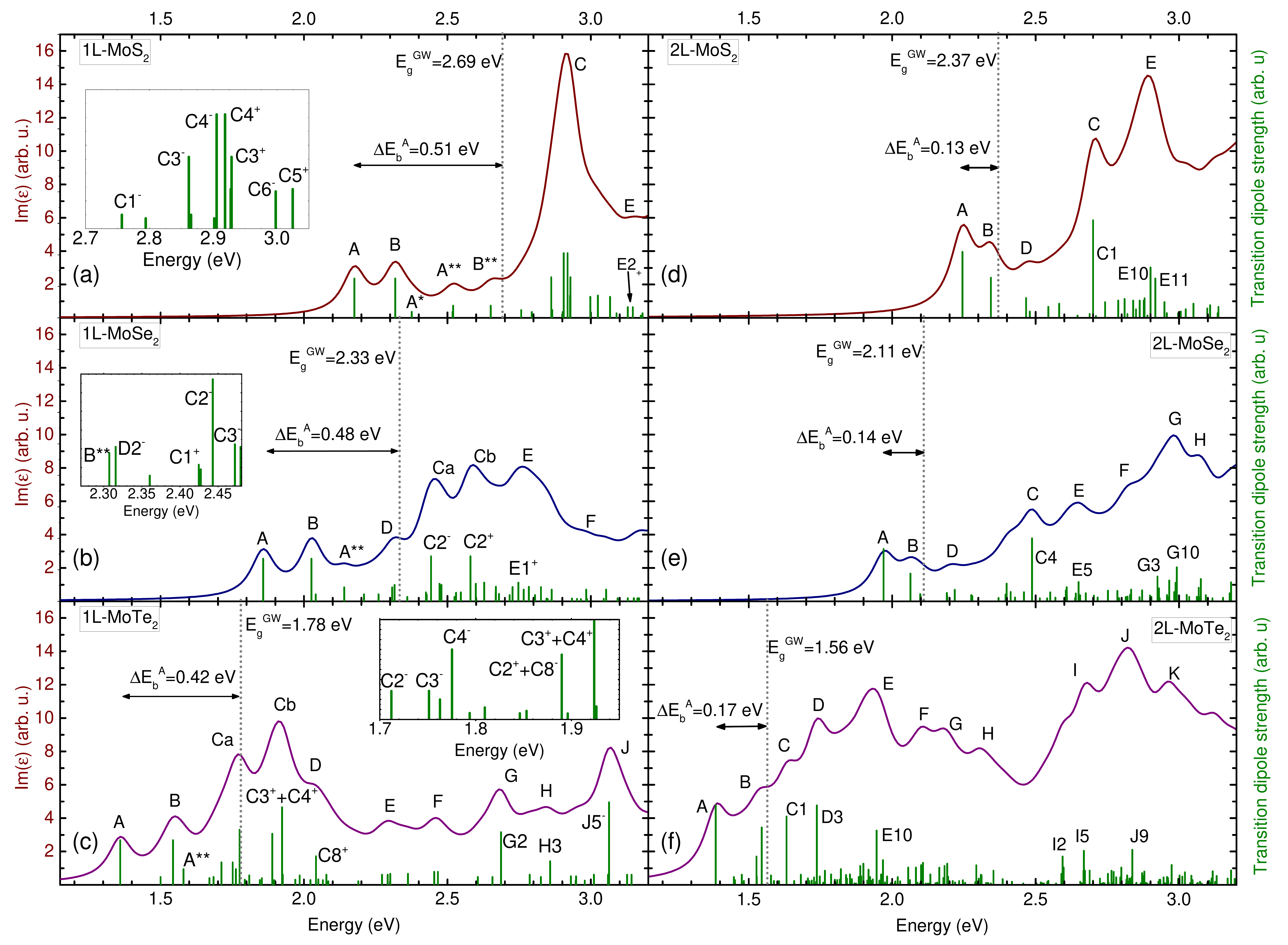}
\caption{\label{fig:absorption} (Color online) Calculated imaginary parts of the dielectric function for (a)-(c) mono- and (d)-(f) bilayer molybdenum and decomposition into the main excitonic transitions. Electron-hole effects were included by solution of the Bethe-Salpeter Equation; spin-orbit splitting was included \emph{a posteriori} for the monolayer systems. With the exception of \textsl{A} and \textsl{B} and their excitations, the excitonic transitions were labeled after the peak in the dielectric function they contribute to. Transitions with the same label in different materials might thus not correspond. The added '-' and '+' mark transitions of same nature that are split by spin-orbit interaction. We broadened the peaks by a Lorentzian of width 0.05\,eV.}
\end{figure*}
\begin{figure}
\centering
\includegraphics*[width=0.9\columnwidth]{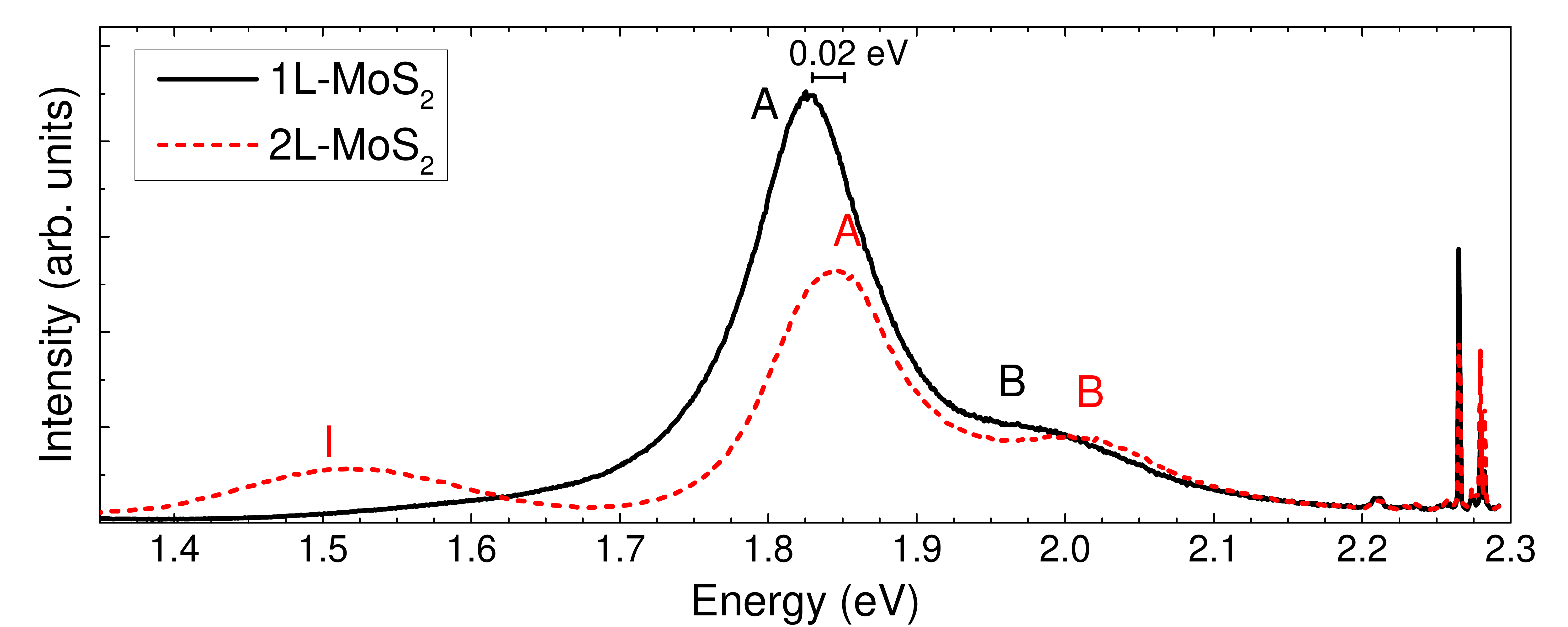}
\caption{\label{fig:expspectrum} (Color online) Photoluminescence spectra of mono- and bilayer MoS$_2$ on Si/SiO$_2$ substrates at an excitation energy of 2.33\,eV. The sharp peaks just below 2.3\,eV are the Raman modes. The shown experimental data has been published previously in  Ref.~\onlinecite{scheuschner-2014}. }
\end{figure}
\section{Computational approach}
We calculated the groundstate properties of all considered materials from density functional theory (DFT) within the common Perdew-Burke-Ernzerhof (PBE) approximation. A 13x13x1 k-point sampling and normconserving pseudopotentials, including semicore states for molydenum, together with a cutoff energy of 1200\,eV were employed on this stage. The DFT electronic bandstructure and wavefunctions served as input for the solution of the excitonic Bethe-Salpeter Equation (BSE)\,\cite{bgw-2} on discrete 30x30 k-point samplings of the Brillouin zone to obtain the dielectric functions including electron-hole effects for the studied dichalcogenides. A sufficient amount of valence and conduction bands to converge the derived optical spectra was included in the calculations. The electronic eigenvalues from DFT were shifted by oneshot G$_0$W$_0$ quasiparticle corrections, using 24x24 k-point grids, a cutoff energy of 300\,eV, and 500/800 bands for the monolayer/bilayer systems.
The dielectric functions and bandstructures of the monolayer systems were corrected {\it a posteriori} for spin-orbit effects following the approach in Ref.~\onlinecite{qiu-2013}. Supercells of 20x20x1 unit cells were used for the plots of the excitonic wavefunctions in real space and the position of the hole was fixed at the Mo atoms in the center of this unit cell. Further details and parameters for trilayer MoS$_2$ can be found in the supplementary material.

\section{Results and Discussion}
\subsection{Absorption properties}
\begin{figure*}
\centering
\includegraphics*[width=\textwidth]{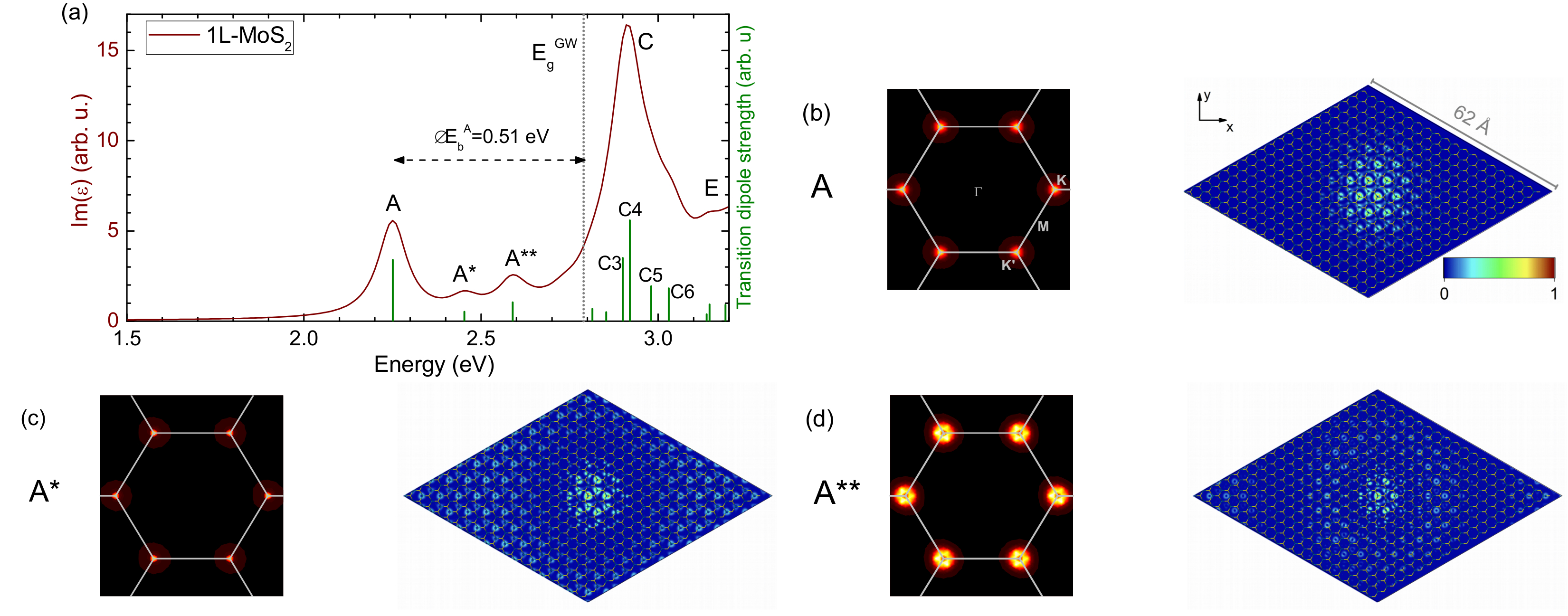}
\caption{\label{fig:MoS2-1L-A} (Color online) (a) Calculated imaginary part of the dielectric function of MoS$_2$ without spin-orbit interaction. (b) Representation of the \textit{A} transition in reciprocal and slice of the excitonic wavefunction in real space ($\|\psi(\vec{r})\|^2$) through the plane of Mo atoms. We used a supercell of 20x20x1 unit cells for the expansion of the excitonic wavefunctions in real space. The position of the hole was fixed at the Mo atom in the center of this supercell. (c) and (d) Same for the first (\textsl{A$^{*}$}) and second excitation (\textsl{A$^{**}$}) of \textsl{A}. The representations of the \textsl{B}, \textsl{B$^{*}$}, \textsl{B$^{**}$} transitions are equal to those of their \textsl{A} counter-parts by construction. }
\end{figure*}

While a number of groups have studied MoS$_2$ in both its mono- and bilayer forms, simulations that include electron-hole effects are to date scarce for MoSe$_2$\,\cite{Ramasubramian-excitons,komsa-excitons,ugeda-MoSe2-excitons} and MoTe$_2$\,\cite{Ramasubramian-excitons,yang-MoTe2-exciton-binding}. 
Figure~\ref{fig:absorption} shows the obtained imaginary parts of the dielectric functions from our calculations. The spectrum of 1L-MoS$_2$, shown in Fig.~\ref{fig:absorption}~(a) agrees well with recent reports\,\cite{wirtz-MoS2-excitons,qiu-2013} in both shape and energies supports our approach for the other materials as well. Inclusion of electron-hole effects induces a number of prominent excitonic peaks that dominate the low-energy imaginary part of the dielectric function: The direct band gap at the \textit{K} point in the Brillouin zone gives rise to two peaks, \textsl{A} and \textsl{B}, that are split from the local spin-orbit coupling and form the fundamental transitions in experimental optical spectra, see Fig.~\ref{fig:expspectrum}. We calculate a binding energy of the \textsl{A} exciton of 0.51\,eV relative to the fundamental electronic band gap of 2.69\,eV from our G$_0$W$_0$ calculations, which fits well into the range of reported values for the \textsl{A} exciton of 0.2-0.57\,eV from experiments\,\cite{MoS2-exciton-binding-1,MoS2-exciton-binding-2,MoS2-exciton-binding-3} and to recent theoretical calculations\,\cite{qiu-erratum,gunnar-MoS2,komsa-excitons}. Apart from the transition forming the absorption onset, the dielectric function features a prominent peak at an energy of 2.9\,eV, which has been previously linked to the prominent '\textsl{C}' feature in photoluminescence spectra at 2.8\,eV\,\cite{C-exc} and consists of a number of strong optical excitations that we will analyse in Sec.~\ref{sec:decomposition}.

While the obtained binding energies of the \textsl{A} (and \textsl{B}) excitons show a good agreement with experiment, our calculations systematically overestimate the absolute peak positions, a fact that our calculations share with previous reports. We attribute parts of this overestimation to the employed G$_0$W$_0$ approximation (we refer to the supplementary material for details) and the neglect of temperature effects in our calculations. The latter have been shown recently\,\cite{mol-sanchez-exciton-temperature} to red-shift the predicted A and B peak positions in MoS$_2$ by about 0.1\,eV.

The dielectric functions of 1L-MoSe$_2$ and 1L-MoTe$_2$ appear qualitatively similar to that of 1L-MoS$_2$, but show more structure, with the main '\textsl{C}' peak being split into a number of subpeaks of similar height, \textsl{Ca} and \textsl{Cb}. For MoSe$_2$, these two dominant subpeaks appear at energies of 2.45\,eV and 2.55\,eV, above the electronic band gap, while in case of MoTe$_2$ the \textsl{Ca} transition moves below the electronic band gap. This is accompanied by a decrease of the fundamental (direct) band gap along the series S$\rightarrow$Se$\rightarrow$Te due to the increase of the in-plane lattice constants. This increase weakens the hybridization of Mo $d$ orbitals with chalcogen $p$ orbitals that is responsible for opening of the fundamental band gap\,\cite{kang-offsets}. Correspondingly, the optical band gap decreases along the series as well.

Our G$_0$W$_0$ calculations predict a direct band gap of 2.33\,eV for 1L-MoSe$_2$, resulting in a binding energy of the \textsl{A} exciton of 0.48\,eV, which is slightly decreased compared to MoS$_2$ and in good agreement with recent studies based on Keldysh electrostatic potentials\,\cite{komsa-2015, kidd-2016}.
The \textsl{A} peak appears at an energy of 1.86\,eV with a spin-orbit-splitting of the \textsl{B} peak by 166\,meV. These values can be compared to the recent report by Ugeda \emph{et. al}\,\cite{ugeda-MoSe2-excitons}, who studied layers of MoSe$_2$ on bilayer graphene by STS and photoluminescence spectroscopy and BSE@GW calculations. Experimentally, they obtained an electronic band gap of E$_g$=2.18\,eV, and an exciton binding energy of E$_b^A$=0.55\,eV, which places the \textsl{A}-peak position at 1.63\,eV. As for MoS$_2$, our calculated exciton binding energy is thus close to experimental values, while the error in the predicted peak position appears to be contained in an overestimated G$_0$W$_0$ band gap compared to experiment. On the other hand, the theoretical calculations in\,\cite{ugeda-MoSe2-excitons} yielded a larger exciton binding energy of 0.63\,eV, which compensated for the overestimated band gap (E$_g$=2.26\,eV) and restored the experimental peak position. Substrate screening was identified to be the main cause of error between theoretical and experimental values in that case and we expect this to add to previously mentioned inaccuracies due to coarse \textit{k}-point samplings.

For MoTe$_2$, we obtain a G$_0$W$_0$ band gap of 1.78\,eV and an \textsl{A} peak at 1.36\,eV, which, again, is about 0.2\,eV higher than the \textsl{A}-peak energy of 1.1\,eV from photoluminescence experiments reported in a recent study\,\cite{yang-MoTe2-exciton-binding}. While the theoretical calculations in\,\cite{yang-MoTe2-exciton-binding} yielded a G$_0$W$_0$ band gap very similar to ours, they predicted a significantly higher exciton binding energy of about 0.6\,eV that led to a good agreement of the simulated and predicted A peak energies and resembles the earlier results by Komsa \emph{et. al}\,\cite{komsa-excitons}. The available theoretical and experimental data thus suggest a binding energy of the A exciton between 0.4 and 0.65\,eV.

For the bilayer materials, it is well established that stacking two layers of MoS$_2$ leads to a transition from direct to indirect semiconductor due to the additional interlayer interactions. At the same time, the direct electronic band gap decreases relative to the monolayer forms due to weaker quantum confinement of the electronic wavefunctions. For bilayer MoS$_2$, our G$_0$W$_0$ calculations predict a reduction of the direct band gap to a value of 2.37\,eV. 
We note that we did not include corrections for spin-orbit interactions in the bilayer systems. This causes an underestimation of the splitting of the \textsl{A} and \textsl{B} peaks in our calculations, which arises solely from band splitting by virtue of interlayer interactions of 100\,meV and introduces an additional error on the band gap of order 50\,meV.
On the other hand, the binding energy of the \textsl{A} peak significantly decreases compared to monolayer MoS$_2$ to a value of 0.13\,eV. This places the \textsl{A} peak at an energy of 2.24\,eV, \emph{i.e.} slightly blue-shifted compared to the \textsl{A} peak in monolayer MoS$_2$, compare to the simulated absorption spectrum of monolayer MoS$_2$ in absence of spin-orbit corrections in Fig.~\ref{fig:MoS2-1L-A}. This is in good agreement with a recent theoretical study\,\cite{palumno-2015} and matches the slight blueshift in the experimental spectra in Fig.~\ref{fig:expspectrum}. On the other hand, the relative positions of the \textsl{A} peaks of mono- and bilayer systems appears to depend on the sample and both spectra with blue-shifted\,\cite{splendiani-2010,scheuschner-2014,MoS2-tunable-PL} and red-shifted\,\cite{mak-2010} bilayer \textsl{A} peak have been reported.
We attribute the reduced binding energies of the \textsl{A} exciton to the proximity of the adjacent layer, which increases the screening of the Coulomb interaction between the electron-hole pair.

At the same time, changes in the electronic structure due to the second layer split the high \textsl{C} peak from monolayer MoS$_2$ into two dominant peaks, \textsl{C} and \textsl{E}, at energies of 2.7 and 2.9\,eV in 2L-MoS$_2$. We note that our calculations can only account for direct transitions and thus do not include the low-energy peak '\textsl{I}' that is observed in photoluminescence spectra due to the fundamental indirect band gap in few-layer MoS$_2$, refer to Fig.~\ref{fig:expspectrum}. Inclusion of such indirect transitions would lead to a shift of oscillation strength from the \textsl{A} exciton to the '\textsl{I}' transition compared to our simulations. 

We find similar results for 2L-MoSe$_2$ and 2L-MoTe$_2$. The binding energy of the \textsl{A} peak of 2L-MoSe$_2$ greatly decreases to a value of E$_b^A$=0.14\,eV compared to the electronic band gap of E$_g$=2.11\,eV and the A peak position of 1.97\,eV is slightly above that of the monolayer form. The dominant \textsl{C} peaks from 1L-MoSe$_2$ contain a significantly lower oscillation strength in 2L-MoSe$_2$ in our calculations, while most of the spectral weight is shifted to a tall peak at higher energies. In 2L-MoTe$_2$, the binding energy of the \textsl{A} exciton of E$_b^A$=0.17\,eV continues the trend and is slightly larger than for the other two bilayer materials. A possible reason might be the increasing layer separation 
along the series 2L-MoX$_2$ with X=S,Se,Te, that is mirrored in the increase of out-of-plane lattice constants in the bulk materials. The larger layer separation might weaken the effect of an adjacent layer on the screening that counteracts and cancels the binding energy trend observed for the monolayers. 
In the following sections we will have a closer look on the detailed compositions of the peaks in the calculated dielectric functions.

\subsection{Decomposition of excitonic transitions in monolayer MoS$_2$}\label{sec:decomposition}
For the sake of clarity, we will start with the dielectric function in the absence of spin-orbit effects, as shown in Fig.~\ref{fig:MoS2-1L-A}~(a). In agreement with expectations, we find that the \textsl{A} peak is caused by excitations from the valence band maxima at the \textit{K} and \textit{K'} points in the Brillouin zone to the lowest conduction band, see Fig.~\ref{fig:MoS2-1L-A}~(b). 
In real space, this strong confinement to the \textit{K} point manifests in a relatively large Bohr radius of the exciton wavefunction of about 16\,\AA, if we define the Bohr radius as full-width-half-maximum (FWHM) of the upper envolope and place the hole on the central Mo atom in the supercell. The excited electrons are mainly in $p_x$ and $p_y$ orbitals, which dominate the conduction band edge around \textit{K} and confine the exciton wavefunction to the MoS$_2$ plane.

%A'/B' and D/D': 
A number of additional transitions with lower dipole-strength appear in the energy range of 2.45-2.55 eV. Without inclusion of spin-orbit coupling, these transitions give rise to two additional features between the \textsl{A} and \textsl{C} peaks, see Fig.~\ref{fig:MoS2-1L-A}~(a). The first of these transitions is strongly localized in reciprocal space and corresponds to a transition between the valence band maximum and conduction band minimum at the \textit{K} point, suggesting that it is indeed the first excited state of the \textsl{A} transition, \textsl{A$^{*}$}. This is confirmed by a node in the real-space representation of the excitonic wavefunction [plotted in Fig.~\ref{fig:MoS2-1L-A}~(c)] at a distance of 23.5\,\AA\space from the hole. 
The second peak corresponds to transitions between the band edges very close to the \textit{K}  point and is more delocalized in reciprocal space compared to \textsl{A} and \textsl{A$^{*}$}. Our analysis reveals two nodes in the excitonic wavefunction at distances of roughly 11.5\,\AA\space and 34\,\AA\space from the hole, suggesting that this transition is the second excited state, \textsl{A$^{**}$}, of the \textsl{A} peak. 
\begin{figure}
\centering
\includegraphics*[width=\columnwidth]{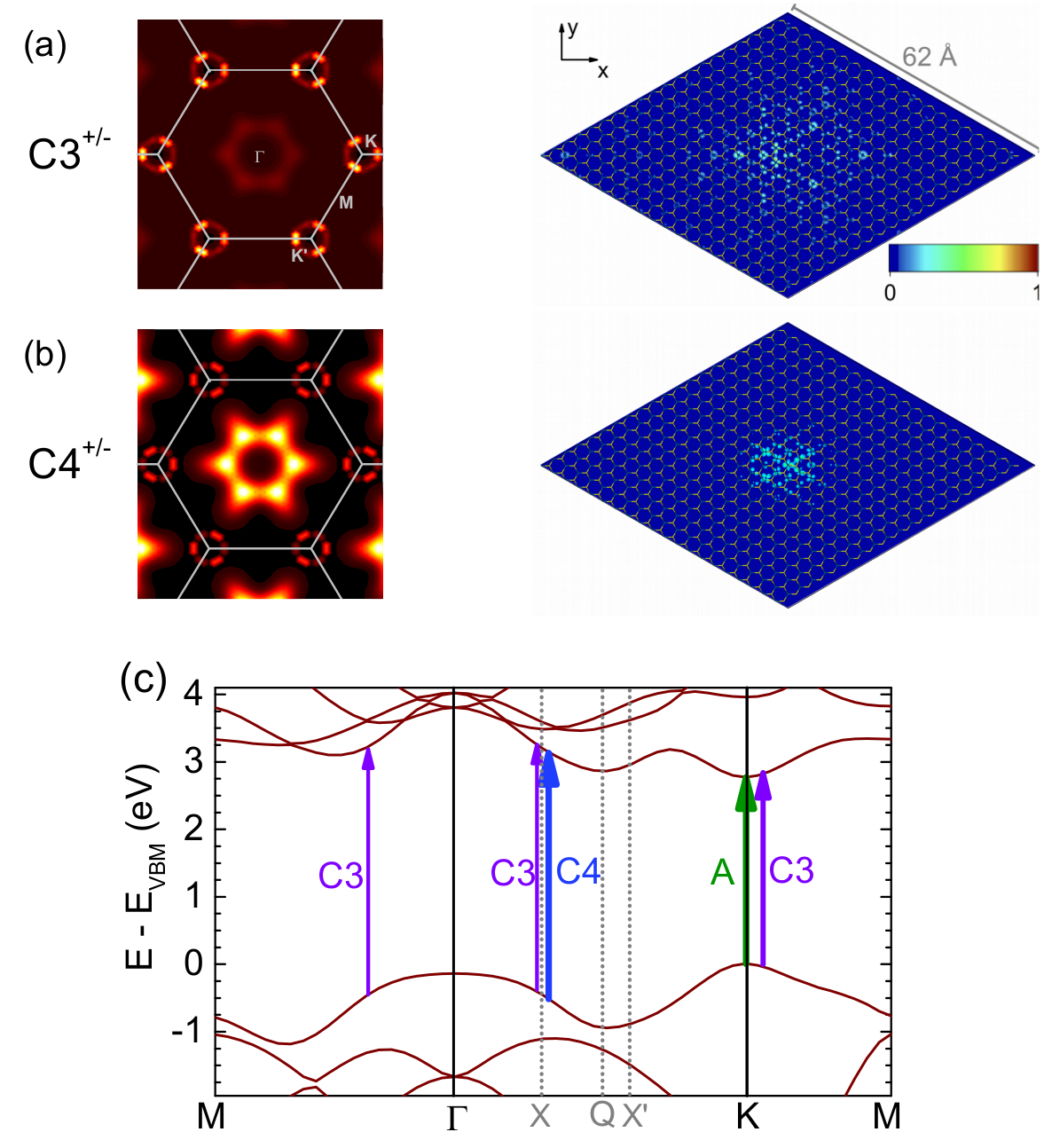}
\caption{\label{fig:MoS2-1L-C} (Color online) Reciprocal and real space wavefunctions of (a) the \textsl{C3$^{+/-}$} and (b) the \textsl{C4$^{+/-}$} transitions contributing to the \textsl{C} peak in the dielectric function of MoS$_2$. Both dominant contributions and the A transition are indicated in the electronic bandstructure in (c). As in Fig.~\ref{fig:MoS2-1L-A}, we did not include spin-orbit effects for the sake of simplicity.}
\end{figure}
\begin{figure*}

\centering
\includegraphics*[width=\textwidth]{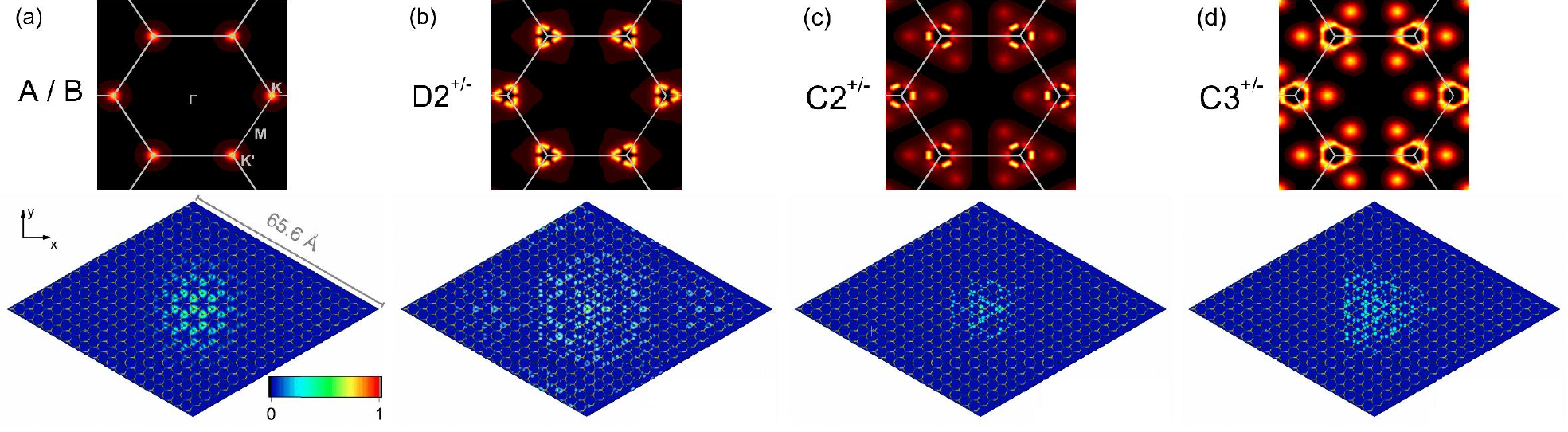}  %(d)

\caption{\label{fig:MoSe2-1L-DCE} (Color online) Reciprocal and real space representations of the (a) \textsl{A}/\textsl{B}, (b) \textsl{D2$^{+/-}$}, (c) \textsl{C2$^{+/-}$} and (d) \textsl{C3$^{+/-}$} transition of MoSe$_2$.}
\end{figure*}

The picture slightly changes if spin-orbit effects are included, see Fig.~\ref{fig:absorption}~(a). The strong localization at the \textit{K} point leads to splitting of \textsl{A}, \textsl{A$^{*}$} and \textsl{A$^{**}$} by 148\,meV into three pairs of transitions, \textsl{A}$\rightarrow$\textsl{A}+\textsl{B}, \textsl{A$^{*}$}$\rightarrow$\textsl{A$^{*}$}+\textsl{B$^{*}$} and \textsl{A$^{**}$}$\rightarrow$\textsl{A$^{**}$}+\textsl{B$^{**}$}. These transitions occur between pairs of valence and conduction bands ($v_1\rightarrow c_1$ and $v_2\rightarrow c_2$) that satisfy the Laporte rule of $\Delta l=\pm 1$, $\Delta s=0$ for the angular momentum quantum number $l$ and the spin quantum number $s$. This has two consequences: \emph{(i)} transitions are only allowed between spin-matched Mo $d$ and S $p$ orbitals, which satisfy the required conditions. \emph{(ii)} The absence of inversion symmetry imposes a reverse order of the spin-orbit split bands at \textit{K'} compared to the \textit{K} point. Hence, the excited electrons possess $s_{K}=\frac{1}{2}$ and $s_{K'}=-\frac{1}{2}$ at the \textit{K} and \textit{K'} points, respectively, in the \textsl{A} exciton, while $s_{K}=-\frac{1}{2}$ and $s_{K'}=\frac{1}{2}$ for the \textsl{B} exciton, giving rise to optical dichroism. 
We further find that the dark \textsl{A$^{*}$} exciton is resonant with the \textsl{B} transition in our calculations, while \textsl{B$^{*}$} is almost degenerate to the considerably brighter \textsl{A$^{**}$} transition. Our calculations thus suggest that the two peaks between \textsl{B} and \textsl{C} in Fig.~\ref{fig:absorption}~(a) do not correspond to \textsl{A$^{*}$} and \textsl{B$^{*}$} but to the spin-orbit split excitations of the \textsl{A} and \textsl{B} transitions, \textsl{A$^{**}$} and \textsl{B$^{**}$}. 

%C:
The \textsl{C} peak is made up of a number of transitions in the range of 2.7-3.0 eV. Two dark contributions, \textsl{C1$^{-}$} and \textsl{C2$^{-}$} (not shown), appear at the lower end of the peak at energies of 2.75 and 2.79\,eV. They correspond to $v_1\rightarrow c_1$ transitions in the vicinity of \textit{K}/\textit{K'}, from points on the \textit{$\Gamma$}-\textit{K} (\textsl{C1$^{-}$}) and \textit{K}/\textit{K'}-\textit{M} lines (\textsl{C2$^{-}$}). %orbitals
The corresponding transitions from $v_2$ to  $c_2$, \textsl{C1$^{+}$} and \textsl{C2$^{+}$}, are shifted to higher energies due to the spin-orbit splitting of the valence band top around the \textit{K} and \textit{K'} points. Approximating the binding energy as the difference between the transition energy and the electronic band gap on the point of the main contribution on the \textit{$\Gamma$}-\textit{K} or \textit{K}-\textit{M} lines yields binding energies on the order of 0.005\,eV, which suggests that both \textsl{C1$^{-/+}$} and \textsl{C2$^{-/+}$} are un- or weakly bound electron-hole pairs. Plots of the representation in reciprocal space and the exciton wavefunction of \textsl{C1$^{+/-}$} can be found in the Supplementary Material.

The bright transitions \textsl{C3} and \textsl{C4} appear at slightly higher energies of 2.86-2.92\,eV and carry almost the entire oscillator strength of the \textsl{C} peak, see Fig.~\ref{fig:absorption}~(a). Despite their energetic similarity, we found significant differences between the two transitions: The reciprocal space representation in Fig.~\ref{fig:MoS2-1L-C}~(a) shows that \textsl{C3} is dominated by contributions from points on the \textit{K}-\textit{M} lines, with weaker, six-fold degenerated contributions around a point \textit{X} at about $\frac{1}{3}$ of the \textit{$\Gamma$}-\textit{K} lines. As for the other transitions, the close vicinity to \textit{K} and \textit{K'} in reciprocal space leads to a split of \textsl{C3} into two transitions \textsl{C3$^{-}$} and \textsl{C3$^{+}$} that are separated by about 0.07\,eV. The corresponding exciton wavefunction, shown in Fig.~\ref{fig:MoS2-1L-C}~(a), inherits a distinct out-of-plane character (not shown) around the hole position from $p_z$ orbitals that strongly contribute to the conduction band at the \textit{X} point, which contrasts the intra-plane nature of the \textsl{A} and \textsl{B} excitons. Interestingly, the probablity of presence of the excited electron exhibits what appears like additional rings surrounding a core with radius on the order
of 20\,\AA, which makes the exciton wavefunction fairly extended. We find similar features for the C1 transitions. It is thus possible that higher order excitations, \emph{e.g.} of \textsl{A}, are mixed into the \textsl{C1} and \textsl{C3} transitions.

In contrast, the second dominant peak, \textsl{C4}, is almost entirely composed of six-fold degenerated contributions at \textit{X}, with relatively minor additional contributions near \textit{K} (and \textit{K'}), see Fig.~\ref{fig:MoS2-1L-C}~(b). The split transitions \textsl{C4$^{-}$} and \textsl{C4$^{+}$} between $v_1$/$v_2$ and $c_1$/$c_2$ are almost degenerate as neither the highest valence nor the lowest conduction bands show a noticeable spin-orbit splitting at \textit{X}, leaving only a relatively minor energy separation of 0.013\,eV from the contributions near \textit{K} and \textit{K'}. The relative expansion of the excitonic transition in reciprocal space leads to a considerable confinement of the exciton wavefunction in real space with a small Bohr radius of 11\,\AA\space compared to that of the \textsl{A} and \textsl{B} excitons. Correspondingly, we find a increased binding energy of about 0.7\,eV compared to the \textit{X} point. These results for \textsl{C4} are in good agreement with previous reports in\,\cite{qiu-2013,qiu-erratum,wirtz-MoS2-excitons}. On the other hand, \textsl{C3} has not been described before, to the best of our knowledge. The similarity of reciprocal space representations and transition energy might suggest that the separation of \textsl{C3} and \textsl{C4} might be an artifact due to symmetry breaking during our calculations. We thus tested the effect of changes in the electronic structure or the sampling in the BSE kernel, but found our results to be robust. 

A number of additional features can be found in an energy window \textless 3\,eV in our calculations. Two additional strong transitions, \textsl{C5$^{+}$} and \textsl{C6$^{+}$}, form a shoulder to \textit{C} at energies around 3\,eV and each have about half the oscillator strength of \textsl{C3$^{+/-}$}. These peaks are composed of unbound electron-hole pairs with wavevectors from a triangular-shaped region around the \textit{K} and \textit{K'} points and vanishing contributions from the entire \textit{$\Gamma$}-\textit{K} line. The spin-orbit splitting pushes \textsl{C5$^{-}$} to an energy of about 2.92\,eV, where it is almost degenerate with the \textsl{C3$^{+}$} transition and thus is expected to be a by-product of the excitation of \textsl{C3$^{+}$}.

The plateau following the \textsl{C} peak [Fig.~\ref{fig:absorption}~(a)] is formed by a number of smaller transitions. Particularly notable are the \textsl{E2$^{+/-}$} transitions that appear at an energy of 3.145\,eV and have the strongest oscillation strength. They consist mainly of contributions within a circle spanned by the six degenerate \textit{X} points and an additional feature at a point '\textit{X'}' [Fig.~\ref{fig:MoS2-1L-C}~(c)] along the \textit{$\Gamma$}-\textit{K} line. Correspondingly, the spin-orbit splitting of \textsl{E2$^{-/+}$} is relatively minor, and on the order of 0.02\,eV. The excitonic wavefunction in real space is well localized within a radius of about 20\,\AA\space and indicates a bound excitonic state. Considering the energy difference to \textsl{C4$^{+/-}$} and the dipole strength, part of \textsl{E2$^{+/-}$} could conceivably be composed of an excited state of \textsl{C4$^{+/-}$} with additional contributions around \textit{$\Gamma$} and on the \textit{$\Gamma$}-\textit{K} lines; however, our calculations do not indicate a node in the excitonic wavefunction. We refer to the supplementary material for the reciprocal and real space representations of \textsl{E2$^{+/-}$}.

\subsection{Excitonic transitions in monolayer MoSe$_2$}
For MoSe$_2$, the \textsl{A} and \textsl{B} transitions and their excitations correspond to those of MoS$_2$ by virtue of the similar electronic dispersion around the \textit{K} and \textit{K'} points, with an energy separation of 168\,meV. The excitonic wavefunction, depicted in Fig.~\ref{fig:MoSe2-1L-DCE}, matches the one of the \textsl{A} exciton of MoS$_2$, with a similar Bohr radius of 17\,\AA. The energy separation between the \textsl{A}/\textsl{B} pair and the \textsl{C} peaks decreases compared to MoS$_2$. As a result, the peak following \textsl{A$^{**}$} [see Fig.~\ref{fig:absorption}~(b)] does not solely originate from \textsl{B$^{**}$} as in MoS$_2$, but has another contribution from the \textsl{D2} peak. This peak corresponds to the \textsl{C1} peak in MoS$_2$ and was pushed to lower energies by the decrease of the band around \textit{K} and \textit{K'}. Fig.~\ref{fig:MoSe2-1L-DCE}~(a) shows our calculated reciprocal space and real space representation of the \textsl{D2} transition. Similar to \textsl{C1} in MoS$_2$, the excitonic wavefunction suggests a weakly or unbound electron-hole pair. 

%C: 
For excitation energies larger than 2.4\,eV, the differences in electronic band structure away from the \textit{K} and \textit{K'} points, induced by exchanging the chalcogen atoms, significantly affect the dielectric function. The dominant contributions come from the \textsl{C2} transitions [Fig.~\ref{fig:absorption}~(b)], which are of similar strength to the \textsl{A} and \textsl{B} peaks. The \textsl{C3} transitions give another strong contribution and appear similar to \textsl{C2}, but are more delocalized in reciprocal space and have about half the oscillation strength of \textsl{C2}, on the same level as the \textsl{A$^{**}$} and \textsl{B$^{**}$} transitions. The plots of reciprocal space representations in Fig.~\ref{fig:MoSe2-1L-DCE}~(b), (c) reveal two main qualitative differences between the dominant \textsl{C} transitions in MoS$_2$ and MoSe$_2$: For one, \textsl{C2} and (particularly) \textsl{C3} in MoSe$_2$ show a significantly stronger weight of contributions in the vicinity of the \textit{K} and \textit{K'} points compared to \textsl{C3} and \textsl{C4} in MoS$_2$, see Figs.~\ref{fig:MoS2-1L-C}~(a),(b) and~\ref{fig:MoSe2-1L-DCE}~(c),(d). This causes a noticeable spin-orbit splitting into subtransitions \textsl{C2$^{+/-}$} and \textsl{C3$^{+/-}$} that give rise to the two peaks \textsl{Ca} and \textsl{Cb} in the dielectric function, with an additional contribution of \textsl{D2$^+$} to \textsl{Ca}. The second difference is found for the contributions away from the \textit{K} and \textit{K'} valleys. While these originate mainly from a point closer to \textit{$\Gamma$} in MoS$_2$, this six-fold degenerated \textit{X} point in MoSe$_2$ appears further along the \textit{$\Gamma$}-\textit{K} line and conincides with the \textit{X'} point in MoS$_2$ [Fig.~\ref{fig:MoS2-1L-C}~(b)]. The excitonic wavefunctions are well localized in real space, with somewhat larger extension than the \textsl{C4} transition in MoS$_2$, We derive Bohr radii of 10\,\AA\space and 19\,\AA\space for the \textsl{C2} and \textsl{C3} transitions, respectively.

%E: 
A band of transitions of roughly equal strength make up the high \textsl{E} peak between 2.7 and 2.9\,eV in the dielectric function, see Fig.~\ref{fig:absorption}~(b). The strongest contribution comes from the \textsl{E1$^{+}$}, with an oscillation strength similar to that of \textsl{D2$^{+/-}$} or \textsl{C3$^{+/-}$}, and with a second transition, \textsl{E2$^{+}$}, is close in energy. The split-off transitions \textsl{E1$^{-}$} and \textsl{E2$^{-}$} contribute to the \textsl{Cb} peak. Both transitions share some similarity to \textsl{C3}, with main contributions at points at around $\frac{1}{2}$\textit{K}-\textit{M} and at the \textit{X} point. The corresponding wavefunctions are relatively extended but well-defined and could thus suggest a bound excitonic state.

\subsection{Excitonic transitions in monolayer MoTe$_2$}
\begin{figure}

\centering
\includegraphics*[width=\columnwidth]{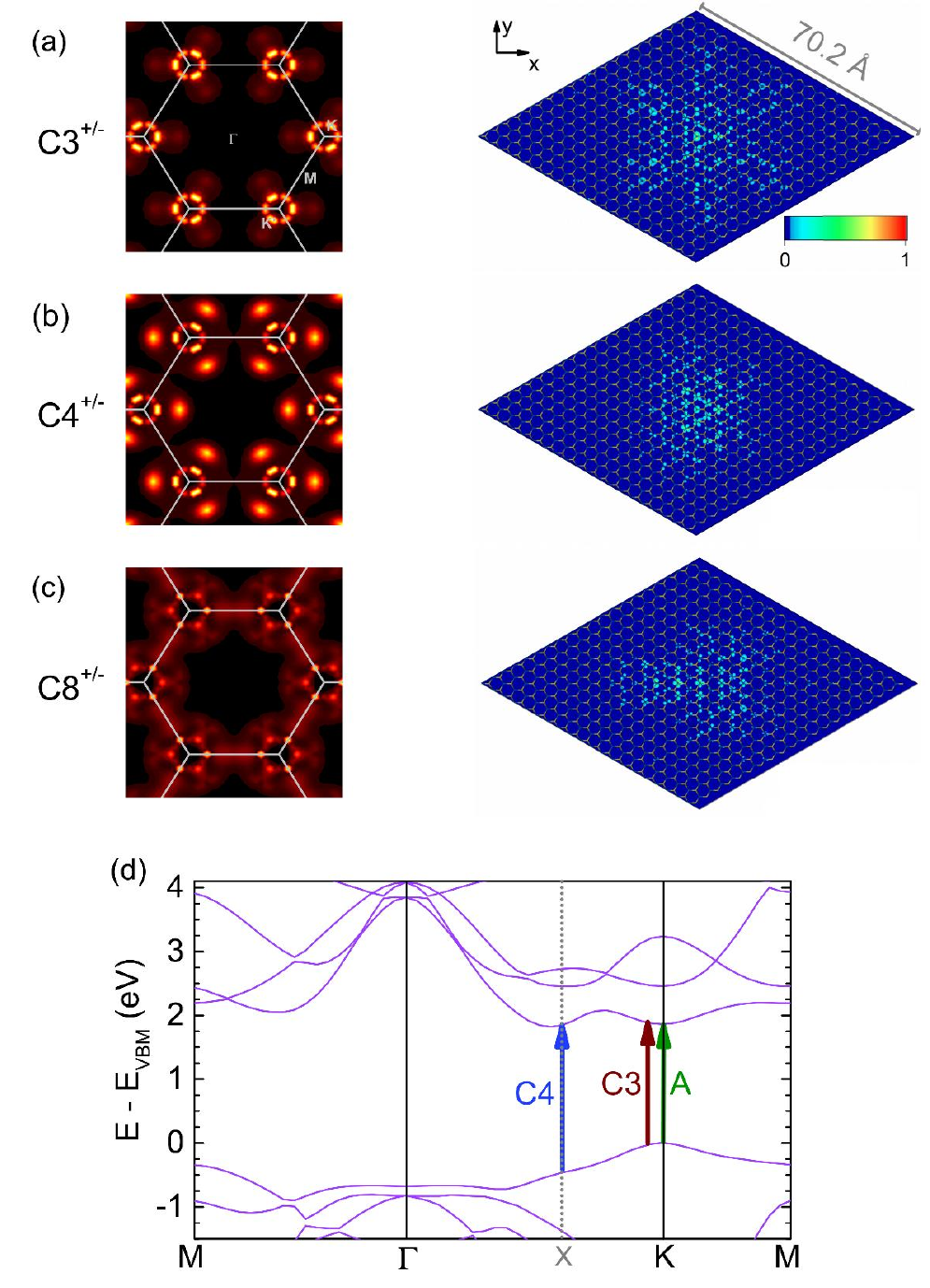}

\caption{\label{fig:MoTe2-1L-C} (Color online) Reciprocal space representations of the (a) \textsl{C3$^{+/-}$}, (b) \textsl{C4$^{+/-}$} and (c) \textsl{C8$^{+/-}$} transitions of MoTe$_2$. (d) Electronic bandstructure of MoTe$_2$ without spin-orbit effects and sketches of the \textsl{A}, \textsl{C3} and \textsl{C4} transitions.}
\end{figure}
In monolayer MoTe$_2$, the spin-orbit splitting at \textit{K} and \textit{K'} of 220\,meV is large enough to push the \textsl{B} transition between the first two excitations of \textsl{A}, see Fig.~\ref{fig:absorption}~(c). The \textsl{B} peak thus consists of a superposition of \textsl{B} with \textsl{A$^{*}$} and \textsl{A$^{**}$}.
The excitonic wavefunction (not shown) assumes the spherical shape seen for MoS$_2$ and MoSe$_2$ and we obtain a slightly increased Bohr radius of about 17\,\AA from the FWHM.

%C  : 
As for MoSe$_2$, the \textsl{C} peak is split into two subpeaks \textsl{Ca} and \textsl{Cb}. \textsl{Ca} mainly consists of the \textsl{C4$^{-}$} transition with smaller contributions from \textsl{C2$^{-}$} and \textsl{C3$^{-}$}. Analysis of the reciprocal space representations show that \textsl{C2$^{+/-}$} correspond to the \textsl{D2$^{+/-}$} transitions in MoSe$_2$, while \textsl{C3} and \textsl{C4} show strong similarities to the \textsl{C2$^{+/-}$} and \textsl{C3$^{+/-}$} excitons in MoSe$_2$, see Fig.~\ref{fig:MoTe2-1L-C}~(a). Transitions at the \textit{X} point on the \textit{$\Gamma$}-\textit{K} line contribute less to the overall oscillation strength of the exciton compared to the other materials. We show in Sec.~\ref{sec:nesting} that the evolution of the \textit{X} point position and its relative contributions along the chalcogen series can be understood from band nesting effects in the electronic band structure. As for the other materials, the excitonic wavefunctions of the \textsl{C3$^{+/-}$} and \textsl{C4$^{+/-}$} transitions are well localized in real space, but have a larger extension

While \textsl{Ca} consists of well-separated excitonic transitions, the composition of \textsl{Cb} appears to be more complex. We find that the \textsl{Cb} peak is almost completely made up of two bright sub-peaks of large oscillation strength: The higher subpeak at 1.92\,eV is formed by a superposition of the \textsl{C3$^{+}$} and \textsl{C4$^{-}$} transitions with an energy separation of only 0.001\,eV. This originates in the smaller energy separation between \textsl{C4$^{-}$} and \textsl{C4$^{+}$} of 150\,meV due to the significantly larger contribution at the \textit{X} point, where the spin-orbit splitting of $v_1$/$v_2$ and  $c_1$/ $c_2$ is smaller than at \textit{K}.  

The second, weaker, sub-peak at 1.89\,eV consists of the \textsl{C2$^{+}$} and \textsl{C8$^{-}$} transitions, which are degenerate ($\Delta E_b=0.0005$\,eV) in our calculations. As Fig.~\ref{fig:MoTe2-1L-C}~(b) shows, \textsl{C8$^{-}$} introduces contributions outside of the vicinity of the \textit{K} and \textit{K'} points, particularly from transitions from $v_1$ and $v_2$ to  $c_1$ along the \textit{K}-\textit{M} line. The boundary condition E$_u$(k)=E$_u$(-k) at \textit{M} acts similarly to inversion symmetry and, together with time-reversal symmetry E$_u$(k)=E$_d$(-k), induces a Kramer's degeneracy on the one-electron states of different spins, thus lifting the spin-orbit splitting of the valence band maximum. In a similar fashion as for the higher peak, this leads to a slightly decreased energy separation of \textsl{C8$^{-}$} and \textsl{C8$^{+}$} ($\Delta$E=150\,meV). The latter forms the \textsl{D} shoulder in the dielectric function in Fig.~\ref{fig:absorption}~(c).
Additional prominent poles in the dielectric function appear at higher energies. We refer to the supplementary material for details.

\subsection{Excitonic states in few-layer TMDs}
\begin{figure*}
\centering
\includegraphics*[width=\textwidth]{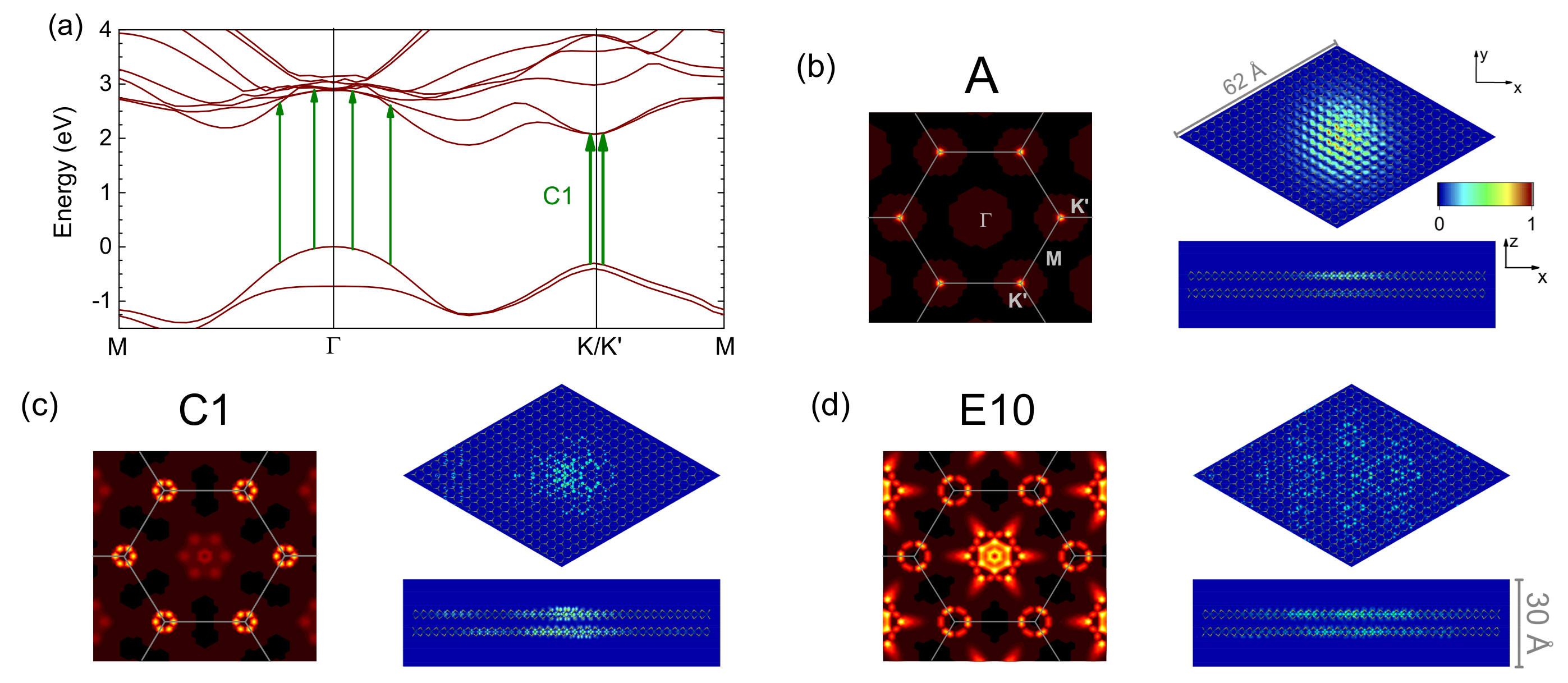}
\caption{\label{fig:MoS2-2L} (Color online) (a) Electronic bandstructure of bilayer MoS$_2$ in the absence of spin-orbit coupling. The reciprocal space representations are plotted for the (b) \textsl{A}, (c) \textsl{C1} and (d) \textsl{E10} transitions. As for the monolayer systems, the excitonic wavefunctions were expanded in real space on a 20x20 supercell. For the sake of clarity, the corresponding real space wavefunctions are shown as slices through the excited layer (upper subpanel) and projected onto the $x$-$z$ plane (lower subpanel). The hole was placed on the center molybdenum atom in the upper MoS$_2$ layer.}
\end{figure*}
Due to the weak interlayer coupling compared to intralayer bonding in layered TMDs, one would expect a relatively small effect of the material thickness on the qualitative nature of the direct excitations. A significant effect here comes from the previously mentioned band splitting due to interlayer interactions, particularly at the \textit{$\Gamma$} and \textit{Q} points that causes a direct-to-indirect bandgap transitions for 1L$\rightarrow$2L, see Fig.~\ref{fig:MoS2-2L}~(a). In bilayer structures (or any even layer number), the added inversion symmetry due to the AB stacking reverses the order of spin-orbit split bands between layers (not included in our presented calculations) and lifts the spin-orbit splitting of the peaks at the absorption onset. In contrast to the monolayer structure, the \textsl{A} peak seen in experiments is thus caused by transitions from two degenerate bands of different total angular momentum at the valence band top at \textit{K}/\textit{K'} to two degenerate conduction bands, while the \textsl{B} peak originates from the corresponding (degenerate) spin-orbit- and interlayer-split bands.

Figure~\ref{fig:MoS2-2L}~(b) shows the calculated reciprocal space representation of the \textsl{A} peak of bilayer MoS$_2$ and the real space excitonic wavefunctions projected onto the x-y and x-z planes, respectively. While the excitonic wavefunction is $s$-like as in monolayer, the Bohr radius of the \textsl{A} transition increases to a value of 20\,\AA\space as a result of the stronger dielectric screening in the bilayer material. 
On the other hand, we find a relatively small spill-over of the excitonic wavefunction into the second layer if an electron-hole pair is excited in the upper layer. This agrees well with recent resonant Raman studies\,\cite{scheuschner-interlayer-modes}, where few-layer MoS$_2$ acts as a superposition of $N$ quasi-independent layers if in resonance with the \textsl{A} and \textsl{B} excitons. The confinement to one plane can be readily explained by the composition of the band edges at the $K$ point, which consist of 
Mo $d_{xy}$, $d_{x^2-y^2}$ (valence band) or Mo $d_{z^2}$ (conduction band) orbitals that are hybridized with sulfur $p_x$ and $p_y$ states and thus extend mainly in-plane. The \textsl{D} peak in the imaginary dielectric function in Fig.~\ref{fig:absorption}~(d) consists of a superposition of \textsl{A$^{**}$} and a second peak of low oscillation strength, \textsl{D1}. Interestingly, this darker second peak is a direct transition at the \textit{$\Gamma$} point, which is energetically closer to the absorption onset due to the strong splitting from to interlayer hybridization of the $p_z$ orbitals that make up the valence band maximum at \textit{$\Gamma$}. It appears from its delocalized excitonic wavefunction that this transition does not form a bound electron-hole pair.

\begin{figure}[tb]
\includegraphics*[width=\columnwidth]{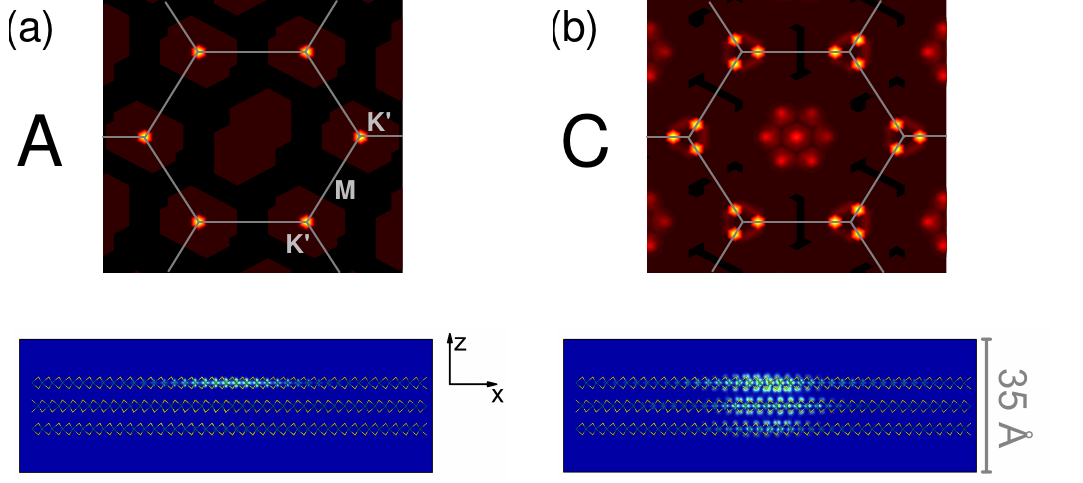}
\caption{\label{fig:MoS2-3L} (Color online) Reciprocal space representations and corresponding excitonic wavefunction in real space (projected onto the the $x$-$z$ plane) of the \textsl{A} and \textsl{C} transitions in trilayer MoS$_2$ corresponding to the \textsl{A} and \textsl{C1} excitons in 2L-MoS$_2$. As for the other materials, the hole was placed on the central Mo atom in the upper-most layer.}
\end{figure}

The strongest single transition in the spectrum is \textsl{C1} that makes up the \textsl{C} peak in the imaginary dielectric function. It appears to correspond to the \textsl{C3} transition in monolayer MoS$_2$, with a considerable contribution away from the vicinity of the \textit{K} and \textit{K'} points. However, the 'off-\textit{K}' contributions appear significantly closer to the \textit{$\Gamma$} point, see Fig.~\ref{fig:MoS2-2L}~(c), which should render the nature of the transitions more \textit{$\Gamma$}-like compared to monolayer MoS$_2$. This $p_z$ character of the conduction band results in a strong interlayer nature of the excitonic wavefunction, which significantly spills over the neighbouring layers if the upper layer is excited. This couples the few-layer structure for optical measurements that are resonant with the \textsl{C} exciton and, \emph{e.g.} activates interlayer resonant Raman modes in MoS$_2$\,\cite{scheuschner-interlayer-modes}. On the other hand, we find that the modified orbital makeup of the \textsl{C} transition changes the shape and the extent of the excitonic wavefunction within the plane of the excited MoS$_2$ layer as well. The interlayer character of the \textsl{C} exciton also carries over to the trilayer system, where the exciton wavefunction is significantly delocalized over all three layers. In contrast, the exciton wavefunction of the \textsl{A} exciton is confined to the excited layer, see Fig.~\ref{fig:MoS2-3L}.

Two additional bright transitions of similar oscillation strength give rise to the dominant \textit{E} peak in the dielectric function of 2L-MoS$_2$ at an energy around 2.9\,eV, see Fig.~\ref{fig:absorption}~(d). While bearing some resemblance in reciprocal space to the \textsl{C1} transition, the contributions are mainly localized within the hexagonal region around the \textit{$\Gamma$} point and on the \textit{$\Gamma$}-\textit{K} line and might correspond to the \textsl{C4} in monolayer MoS$_2$. However, the calculated excitonic wavefunctions appear delocalized within and between the layers, see Fig.~\ref{fig:MoS2-2L}~(d).

Our results for bilayer MoSe$_2$ draw a similar picture as those for MoS$_2$. The strongest transition in our calculations has a similar reciprocal space representation as the \textsl{C3} excitons in the monolayer material and forms the \textsl{C} peak in the imaginary dielectric function [Fig.~\ref{fig:absorption}~(e)]. As for MoS$_2$, we find a noticeable out-of-plane component of the excitonic wavefunction. The main factor is the orbital composition of the involved first valence and conduction bands at the \textit{X} points at around $\frac{3}{5}$ of the \textit{$\Gamma$}-\textit{K} line, which consist of selene p$_y$ and p$_z$ and Mo $d$ states, while the first valence band has a larger Se p$_x$ character. 

Compared to MoS$_2$ and MoSe$_2$, the dielectric function of 2L-MoTe$_2$ features a number of bright transitions that dominate the dielectric function below an energy of 2.0\,eV, see Fig.~\ref{fig:absorption}~(f). The \textsl{C} and \textsl{D} peaks, see Fig.~\ref{fig:absorption}, are dominated by transitions between the two highest valence bands $v_1$ and $v_2$ to the two lowest conduction bands that in reciprocal space show a striking resemblance to \textsl{C3} and \textsl{C4} in the monolayer. Here, the \textsl{C1} transition is mainly localized at the \textit{K} point, with a relatively low contribution at the \textit{X} point away from \textit{K}. The excitonic wavefunction shows a relatively small Bohr radius of 13\,\AA\space within the excited layer, but, as for MoS$_2$ and MoSe$_2$, is delocalized over the bilayer structure. In contrast, the \textit{D3} transition, which dominates the \textsl{D} peak, has no contribution from \textit{K} or its immediate neighbourhood, but mainly consists of transitions at the \textit{X} point. The corresponding excitonic wavefunction appears to be very extended with a radius of more than 25\,\AA, indicating a weakly bound electron-hole pair.

A number of higher features appear in the dielectric functions of both 2L-MoSe$_2$ and 2L-MoTe$_2$ above the bright C transitions and should be accessible through suitable laser energies. The \textsl{E} peak in MoTe$_2$ is predominantly a transition from $v_2$ to the third conduction band  $c_3$ at the \textit{K} (and \textit{K'}) point. The broad \textsl{G} peak in MoSe$_2$ is composed of band transitions that mainly involve the \textit{$\Gamma$} point. Plots of the reciprocal and real space representations of selected transitions of MoSe$_2$ and MoTe$_2$ and of the \textsl{E} transitions in MoS$_2$ can be found in the supplementary information.

\subsection{Band nesting}\label{sec:nesting}
\begin{figure*}
\centering

\includegraphics*[width=\textwidth]{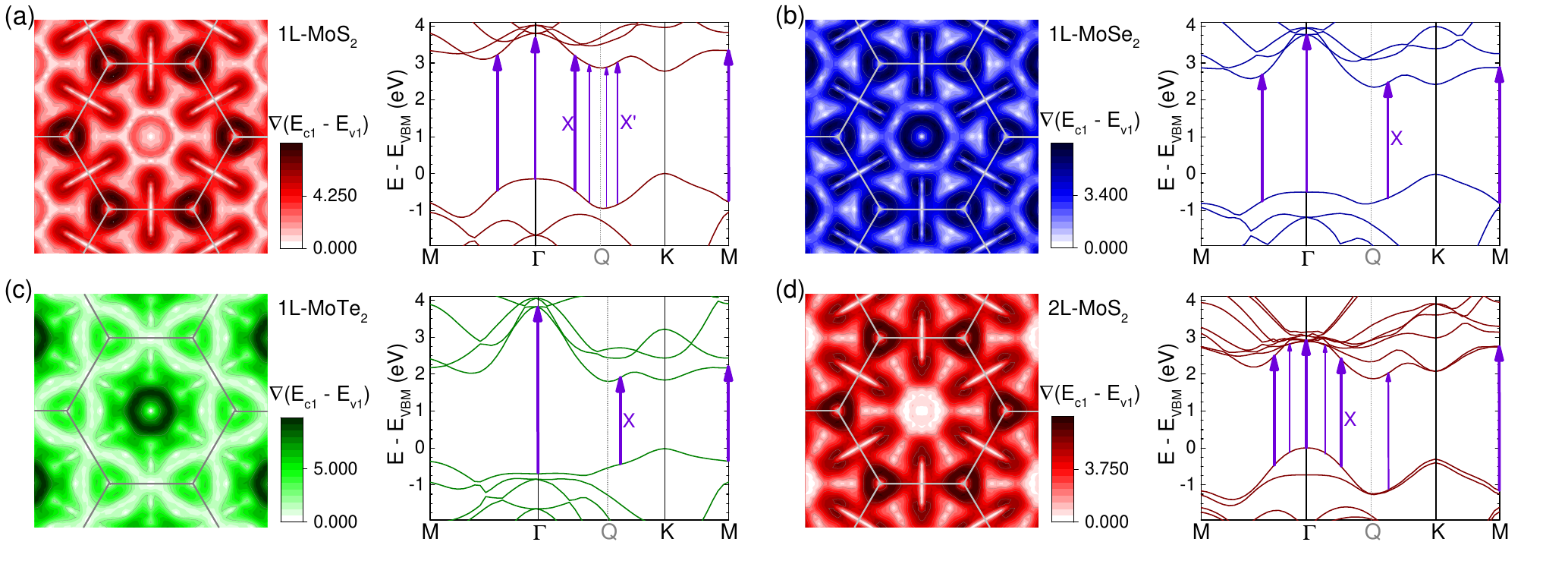}

\caption{\label{fig:nesting} (Color online) Differences of gradient between the highest valence band ($v_1$) and lowest conduction band ( $c_1$) and electronic bandstructures (without spin-orbit coupling) for monolayer (a) MoS$_2$, (b) MoSe$_2$, and (c) MoTe$_2$, and for (d) bilayer MoS$_2$. Arrows in the bandstructures indicate transitions at the band nesting points with large joint density of states.}
\end{figure*}
Following Fermi's Golden Rule, two factors are important for optical absorption of a solid material: One is the dipole matrix element of a specific transition between valence band $v$ and conduction band $c$ at a given \textit{k}-point $k$. The second factor is the joint density of states $\rho(\hbar\omega)=\sum_{v,c}\rho_{vc}(\hbar\omega)$, i.e. the number of different absorption channels for a given energy $\hbar\omega$. The joint density of states between $v$ and $c$ is commonly defined as 
\begin{equation}
\rho_{vc}(\hbar\omega)\propto\int\frac{d\vec{S}(\vec{k})}{ \left|\nabla_k(E_{c}(\vec{k})-E_{v}(\vec{k}))\right|_{E_{c}(\vec{k})-E_{v}(\vec{k})=\hbar\omega} },
\end{equation} 
where integration is performed over a constant energy surface $S$ in reciprocal space\,\cite{carvalho-band-nesting}. It is clear from this definition that major contributions to the joint-density of states come from quasi-stationary regions in reciprocal space where $\left|\nabla_k(E_{c}-E_{v})\right|$ is small or vanishing. On one hand, this is the case at points with a conduction band minimum and valence band maximum, where the individual gradients vanish, i.e. $\left|\nabla_k(E_{c})\right|=\left|\nabla_k(E_{v})\right|=0$ and thus causes strong absorption at the band edges. On the other hand, the condition is also fulfilled at points in the bandstructure, where $\left|\nabla_k(E_{c})\right|\neq 0$ and $\left|\nabla_k(E_{v})\right|\neq 0$, but $\left|\nabla_k(E_{c}-E_{v})\right|\approx 0$. 
Carvalho \emph{et al.}\,\cite{carvalho-band-nesting} recently suggested by DFT calculations that such points of "band nesting" give rise to the dominant peaks in the optical conductivity of various transition metal dichalcogenides.

Figure~\ref{fig:nesting}~(a)-(c) show plots of the gradient of the energy difference between the GW corrected first valence and conduction bands of the three monolayer materials over the first Brillouin zone in the absence of spin-orbit interaction. 
For monolayer MoS$_2$, we find a range of points with small or vanishing gradient in the Brillouin zone, specifically at the high symmetry points and along the \textit{$\Gamma$}-\textit{K} direction. Most notably, the plot reveals a hexagonally shaped area around the \textit{$\Gamma$} point with a low gradient that, when compared to Fig.~\ref{fig:MoS2-1L-C}, fits well to the off-\textit{K} contributions to the \textsl{C3} and \textsl{C4} transitions. Another point of vanishing gradient difference appears at $\frac{2}{3}$ of the \textit{$\Gamma$}-\textit{K} line, which originates from a region close to the \textit{Q} valley in the electronic dispersion, see the bandstructure plot in Fig.~\ref{fig:nesting}~(a), where both conduction and valence band exhibit similar curvature. Due to the band gap of about 3.8\,eV, this region does not noticeably influence the dielectric function up to energies of 3\,eV, where the \textsl{E2} and \textsl{E3} transitions contain first contributions from the band nesting point.

Replacing S with Se has a significant effect on the electronic structure that change the picture, see Fig.~\ref{fig:nesting}~(b). The relative splitting of valence and conduction band at the \textit{$\Gamma$} point increases compared to MoS$_2$, increasing the steepness of the conduction band between \textit{$\Gamma$} and \textit{Q}, and simultaneously pushing the valence band maximum at \textit{$\Gamma$} towards lower energies, thus decreasing the gradient in \textit{$\Gamma$}-\textit{Q} direction. This lifts the band nesting that MoS$_2$ exhibits between \textit{$\Gamma$} and \textit{Q} and increases the transition energy relatively to the fundamental electronic band gap. Correspondingly, the contribution of this point to the low-energy part of the absorption spectra is negligible. This leaves the band nesting point between \textit{Q} and \textit{K}, which is barely affected, as the changes in valence and conduction band dispersion cancel each other. At the same time, the band gap at \textit{Q} decreases relatively to \textit{K} and brings transitions at this point closer to the low-energy part of the absorption spectrum compared to MoS$_2$. The band nesting at this point thus strongly contributes to the dominant \textsl{C2} and \textsl{C3} transitions in MoSe$_2$, compare Fig.~\ref{fig:nesting}~(b) and Fig.~\ref{fig:MoSe2-1L-DCE}~(c),(d).

This trend continues for MoTe$_2$; the splitting at \textit{$\Gamma$} further increases, which counteracts the general lowering of the conduction band compared to the valence band due to the increased interatomic distances and results in a high-gradient difference region around the \textit{$\Gamma$} point that does not contribute to the low-energy absorption. The band nesting point \textit{X} between \textit{Q} and \textit{K} also appears for MoTe$_2$ and strongly contributes to the \textsl{C3} and \textsl{C4} transitions in Fig.~\ref{fig:absorption}~(c). This explains the noticeable similarity in nature of the dominant \textsl{C} transitions in MoTe$_2$ and MoSe$_2$ compared to the \textsl{C3} and \textsl{C4} transitions in MoS$_2$. Further, the band nesting around \textit{X} forms a band of low gradient difference between neighbouring \textit{M} points that surrounds \textit{K} and \textit{K'} and contributes to a relative delocalization of the contributions to excitonic transitions that can, for instance, be seen for the \textsl{C8} transition depicted in Fig.~\ref{fig:MoTe2-1L-C}~(c).

On the other hand, we have seen in the previous section that the interlayer interaction in few-layer materials is another source for modifications in the electronic dispersion compared to their monolayer forms that affects the
quality of optical excitations. In bilayer MoS$_2$, this particularly concerns the region around the \textit{$\Gamma$} point, where hybdridization of the out-of-plane sulfur $p_z$ orbitals between the two layers induces a splitting of the valence band by 0.72 eV. This increases the band energy gradient around \textit{$\Gamma$}. At the same time, we find that the energy difference between the conduction band edges at \textit{$\Gamma$} and \textit{Q} only negligibly changes compared to 1L-MoS$_2$. As
Fig.~\ref{fig:nesting}~(d) shows, this shifts the point of band nesting on
the \textit{$\Gamma$}-\textit{K} lines and forms a compressed region of low gradient difference between valence and conduction band around \textit{$\Gamma$}. The obtained band nesting region fits well to the position of the off-\textit{K} contributions to the \textsl{C} exciton in our calculations and thus suggests that these contributions
are caused by a superposition of transitions with possibly weak oscillation strength that are amplified by the high joint density of states.

\section{Conclusion}
Based on \textit{ab initio} calculations, we showed that the composition and nature of the excitonic peaks in molydenum dichalcogenides is affected by the chalcogen species and interlayer interactions in the system. The bandstructures of the studied mono-and bilayer systems exhibit singularities in the joint-density of states between the first valence and conduction bands that are shifted along the \textit{$\Gamma$}-\textit{K} line as a consequence of changes in the electronic dispersion. This particularly concerns the prominent \textsl{C} peak that has been observed in MoS$_2$ and should also appear in the other materials as well. We predict that the moved band nesting point induces a sizeable spin-orbit splitting in the \textsl{C} peak for MoSe$_2$ and MoTe$_2$ due to its greater vicinity to the \textit{K} point, and a corresponding increasing peak broadening along the series MoS$_2\rightarrow$MoSe$_2\rightarrow$ MoTe$_2$. Further, we confirm the noticeable interlayer character of the excitonic wavefunction of the \textsl{C} peak transition in few-layer MoS$_2$, as opposed to the intralayer character of the fundamental \textsl{A} transition, and show that a similar behaviour can be expected for MoSe$_2$ and MoTe$_2$ as well. This has interesting implications for experimental measurements on N-layer dichalcogenide materials, which thus could be forced to act more decoupled (excitation resonant with \textit{A} exciton) or coupled (excitation resonant with \textsl{C} exciton.)

\section{Acknowledgements}
The authors gratefully acknowledge the North-German Supercomputing Alliance (HLRN) for providing the computational ressources used for the simulations in this work. This work was supported by the European Research Council (ERC) under grant number 259286 and the German Research Foundation (DFG) within SPP1459 "Graphene". The authors thank Nils Scheuschner for discussions and useful input.

%

%%%%% Merge with supplementary material %%%%%%%%%%%%%%%%%

\pagebreak
\clearpage
\pagestyle{plain}
\widetext
\renewcommand{\floatpagefraction}{0.99}
\begin{center}
\textbf{\LARGE Supplementary Materials:}\\
\textbf{\LARGE Light-matter Interactions in Twodimensional Transition Metal Dichalcogenides: 
Dominant Excitonic Transitions in mono- and few-layer MoX$_2$ and Band Nesting}\\
\vspace{0.8cm}
{\Large Roland Gillen and Janina Maultzsch}\\
\vspace{0.2cm}
{Institut f\"ur Festk\"orperphysik, Technische Universit\"at Berlin, Hardenbergstr. 36, 10623 Berlin, Germany}
\end{center}

\setcounter{equation}{0}
\setcounter{figure}{0}
\setcounter{table}{0}
\setcounter{page}{1}
\setcounter{section}{0}
\setcounter{footnote}{0}
\makeatletter
\renewcommand{\theequation}{S\arabic{equation}}
\renewcommand{\thefigure}{S\arabic{figure}}
\renewcommand{\bibnumfmt}[1]{[S#1]}
\renewcommand{\citenumfont}[1]{S#1}

\large

\section{Computational Method}
\begin{figure}[b]
\centering
\includegraphics*[width=0.65\columnwidth]{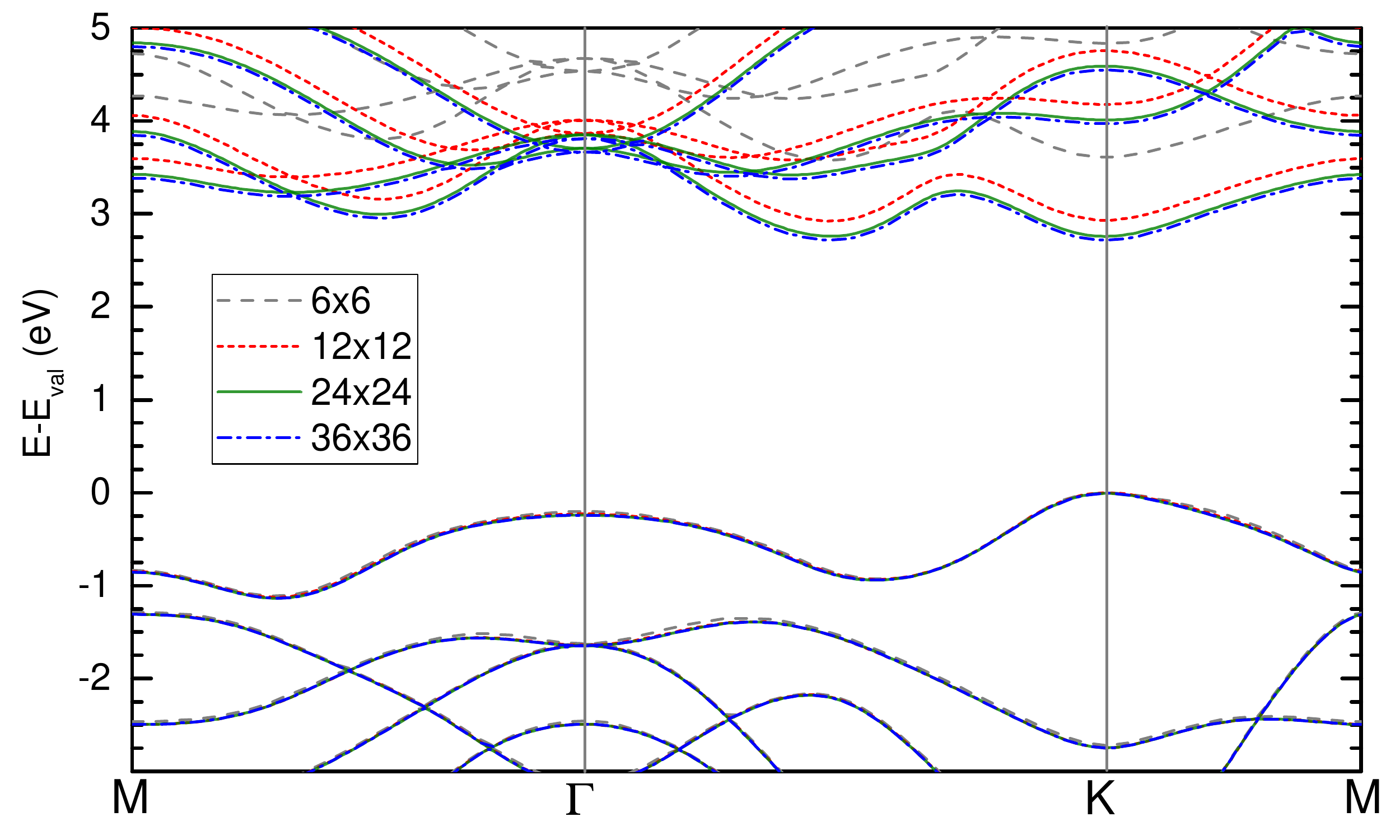}
\caption{\label{fig:GW-MoS2} Simulated electronic bandstructures of monolayer MoS$_2$ from G$_0$W$_0$ calculations and k-point samplings of varying density. The dispersion appears converged for a grid 12x12 points in the quasi-two-dimensional Brillouin zone, while the effect of denser samplings is a rigid relative shift of conduction and valence bands. This can be traced back to the preculiar structure of the static dielectric function $\epsilon_q$(G,G') for small \textbf{q}, refer to\,\cite{huser-MoS2,qiu-arxiv}.}
\end{figure}

The necessary input for the calculations of the optical properties was calculated through density functional theory on the level of the Perdew-Becke-Ernzerhof (PBE) approximation as implemented in the Quantum Espresso package\,\cite{qe}. Using the ATOMIC code, we generated two sets of scalar- and fully relativistic normconserving pseudopotentials for Mo, S, Se and Te and used a cutoff energy of 1200 eV in our calculations. Here, we included the Mo 4s and 4p semi-core electrons in the set of valence electrons, while we found that semi-core states for Se and Te can be neglected without affecting the results. Integrations in reciprocal space were performed on a discrete grid of 13x13x1 points in the Brillouin zone. Using these parameters, we optimized the atomic positions and inplane lattice constants until the interatomic forces and the stress was below thresholds of 0.01\,eV/\AA\space and 0.01\,GPa, respectively, while keeping a vacuum layer of at least 25\,\AA\space in $c$ direction in order to minimize spurious interactions between periodic images. Here, the bi- and trilayer structures were modeled as AB-stacked MoX$_2$ sheets, with inversion symmetry for the bilayer materials. A proper description of the interlayer interaction is crucial\,\cite{scheuschner-2014} in the few-layer structures to obtain the large valence band splitting at the $\Gamma$ point that is observed in ARPES measurements\,\cite{val-split}. For this purpose, we included semi-empirical van-der-Waals corrections from the PBE+D3\,\cite{d3-1,d3-2} scheme, which yields excellent predictions of the in- and out-of-plane lattice constants of layered systems~\footnote{Manuscript in preparation}. Our obtained lattice parameters for the studied mono-, bi- and trilayer systems and the bulk materials are reported in Table~\ref{tab:latconsts}.

We then calculated the optical properties using the BerkeleyGW code\,\cite{hybertsen-louie,bgw-2,bgw-3}. 
The optical spectra including electron-hole interactions were calculated by solving the Bethe-Salpeter Equation (BSE), using 4/6/9 valence bands and 6/12/18 conduction bands for monolayer/bilayer/trilayer systems. The input static dielectric function was calculated using 200/300/350 unoccupied bands for monolayer/bilayer/trilayer systems and an energy cutoff of 300\,eV.
Following the typical approach, the electronic bandstructures from DFT were corrected by G$_0$W$_0$ quasiparticle energies before entering the diagonalization of the BSE kernel matrix. We employed the static remainder technique\,\cite{static-remainder} that allowed us to limit the number of empty bands to 500/800/900 in the monolayer/bilayer/trilayer systems while obtaining sufficiently converged transition energies for our purposes.
The divergence of the Coulomb potential and spurious interactions with neighbouring cells was treated through a Wigner-Seitz truncation scheme\,\cite{ws-truncation}. Using the Hybertsen-Louie plasmon-pole model\,\cite{hybertsen-louie}, we calculated the G$_0$W$_0$ corrections for the monolayer and bilayer systems on a grid of 24x24x1 k-points and a cutoff energy of 300 eV for the screened Coulomb interaction. Based on a series of test calculations in MoS$_2$, see Fig.~\ref{fig:GW-MoS2} and Refs.\,\cite{huser-MoS2,MoSe2-bradley,qiu-arxiv}, we estimate the accuracy of our obtained electronic band gaps to be of order 0.1\,eV. We then used Wannier interpolation to obtain the quasiparticle correction for each band on the k-point grid used for the calculations of the optical spectra.

For better comparability, we chose to employ 30x30x1 k-point grids for the BSE calculations of all studied mono- and bilayer systems, while a 21x21x1 k-point grid and energy cutoffs of 250\,eV were used for both BSE and G$_0$W$_0$ calculations of trilayer MoS$_2$.
Spin-orbit effects were included a posteriori into the spectra and the quasi-particle band structures of the monolayer systems following the scheme reported previously by Louie \emph{et al.}\,\cite{qiu-2013}. The bi- and trilayer systems were not corrected as the \textit{a posteriori} correction does not appear to work very well in these cases due to the degeneracy of the valence band maximum at the $K$ point.
\begin{table}[h!]
\caption{\label{tab:latconsts} Obtained lattice constants and layer separations for all considered systems and the bulk materials.}
\begin{tabular}{ |c|c|c|c| }
\hline
\hline
Material& N layers & lattice constant (in \AA)& layer separation (in \AA)\\
\hline
& 1L & 3.15 & \\
MoS$_2$& 2L & 3.15 & 6.105 \\
& 3L & 3.151 & 6.105 \\
& bulk & 3.153 & 6.08 \\
\hline
& 1L & 3.279 & \\
MoSe$_2$& 2L & 3.28 & 6.45\\
& bulk & 3.284 & 6.435\\
\hline
& 1L & 3.515 & \\
MoTe$_2$& 2L & 3.519 & 6.93\\
& bulk & 3.522 & 6.915\\
\hline
\hline
\end{tabular}
\end{table}

\newpage
\section{Optical spectra}
\begin{figure}[h!]
\includegraphics*[width=\columnwidth]{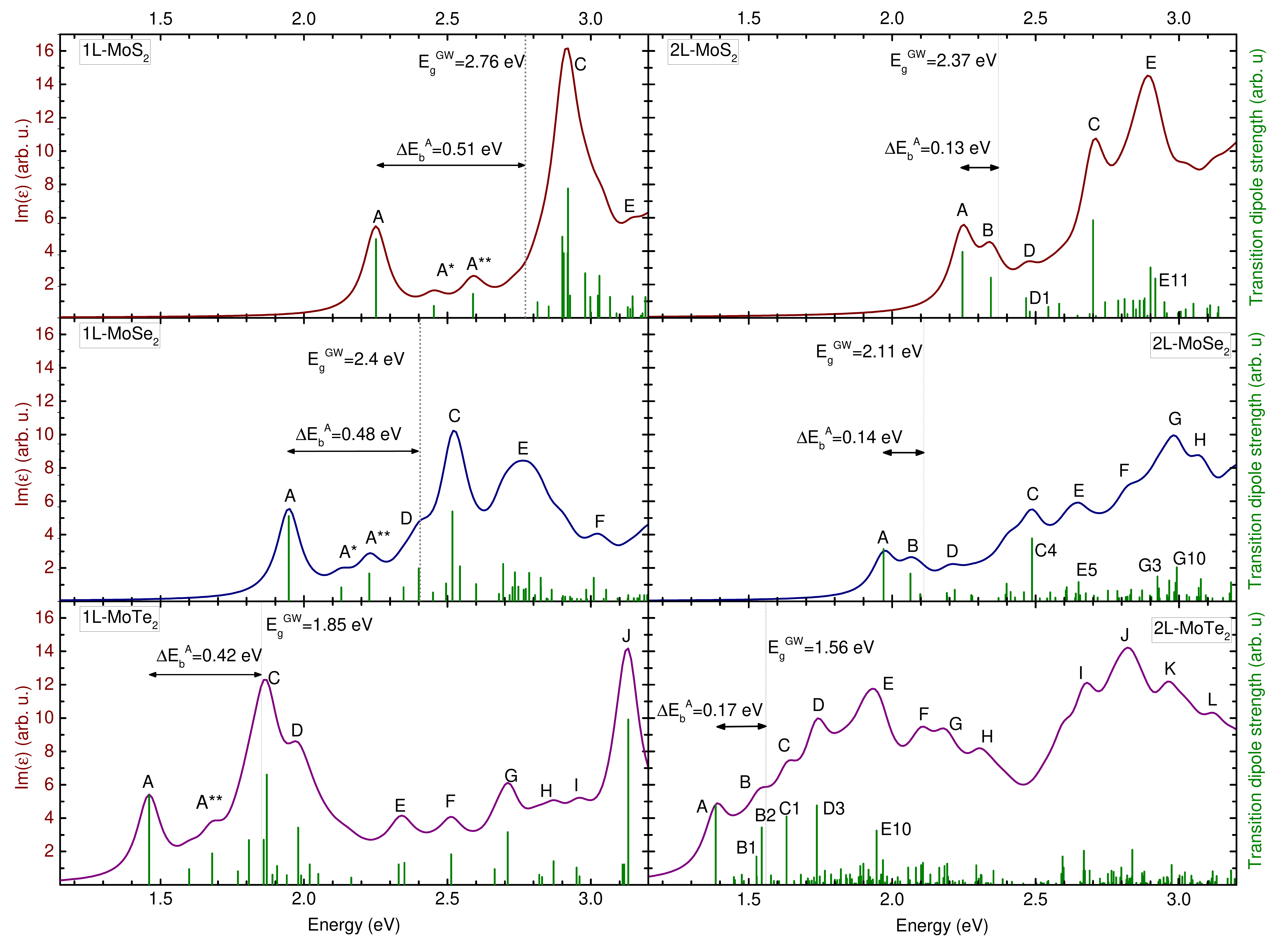}
\captionof{figure}{\label{fig:dielect} Calculated imaginary dielectric functions for mono- and bilayer MoX$_2$ in absence of spin-orbit interaction. As in Fig. 1 of the main text, we broadened the peaks by a Lorentzian of width 0.05\,eV. The decomposition of the peaks into the strongest transitions is shown. E$^{GW}$ is the fundamental (direct) band gap in the calculated quasi-particle bandstructures.}
\end{figure}

%
%\subsection{Dielectric functions without spin-orbit corrections}
%
For comparison with the results in the main text, we here show additional calculated spectra for the six studied mono- and bilayer molybdenum dichalcogenides. Fig.~\ref{fig:dielect} shows the calculated imaginary parts of the dielectric functions without spin-orbit coupling effects for the monolayer systems. On the other hand, we calculated the frequency-dependent absorption coefficient from the obtained dielectric functions (including spin-orbit interaction) and show the resulting simulated absorption spectra in Fig.~\ref{fig:absorption}.

%subsection{Absorbance spectra}
%
\begin{figure}[h!]
\includegraphics*[width=\columnwidth]{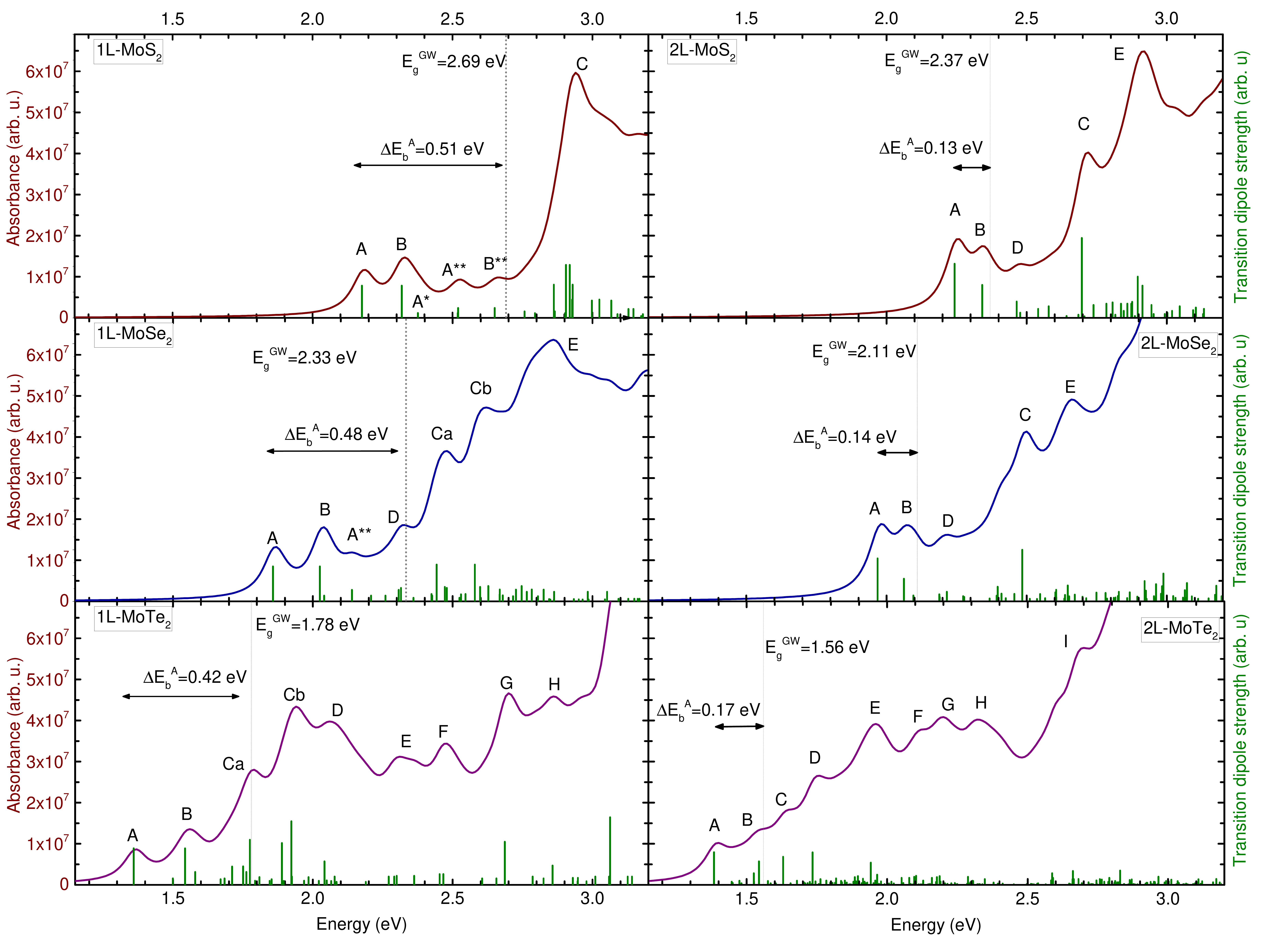}
\caption{\label{fig:absorption} Simulated absorption spectra as derived from the calculated dielectric functions. Spin-orbit interactions were included \textit{a posteriori}. The peaks were broadened by a Lorentzian of width 0.05\,eV.}
\end{figure}

\clearpage
\section{Excitonic wavefunctions in reciprocal and real space}
\subsection{1L-MoS$_2$}
This subsection contains plots of the reciprocal- and real-space representations of additional transitions of monolayer MoS$_2$ contributing to the \textsl{C} and \textsl{E} peaks in the dielectric function, compare to Fig. 1~(a) in the main manuscript, that might be of interest to the reader.
\begin{figure}[hb!]
\centering
\includegraphics*[width=0.93\columnwidth]{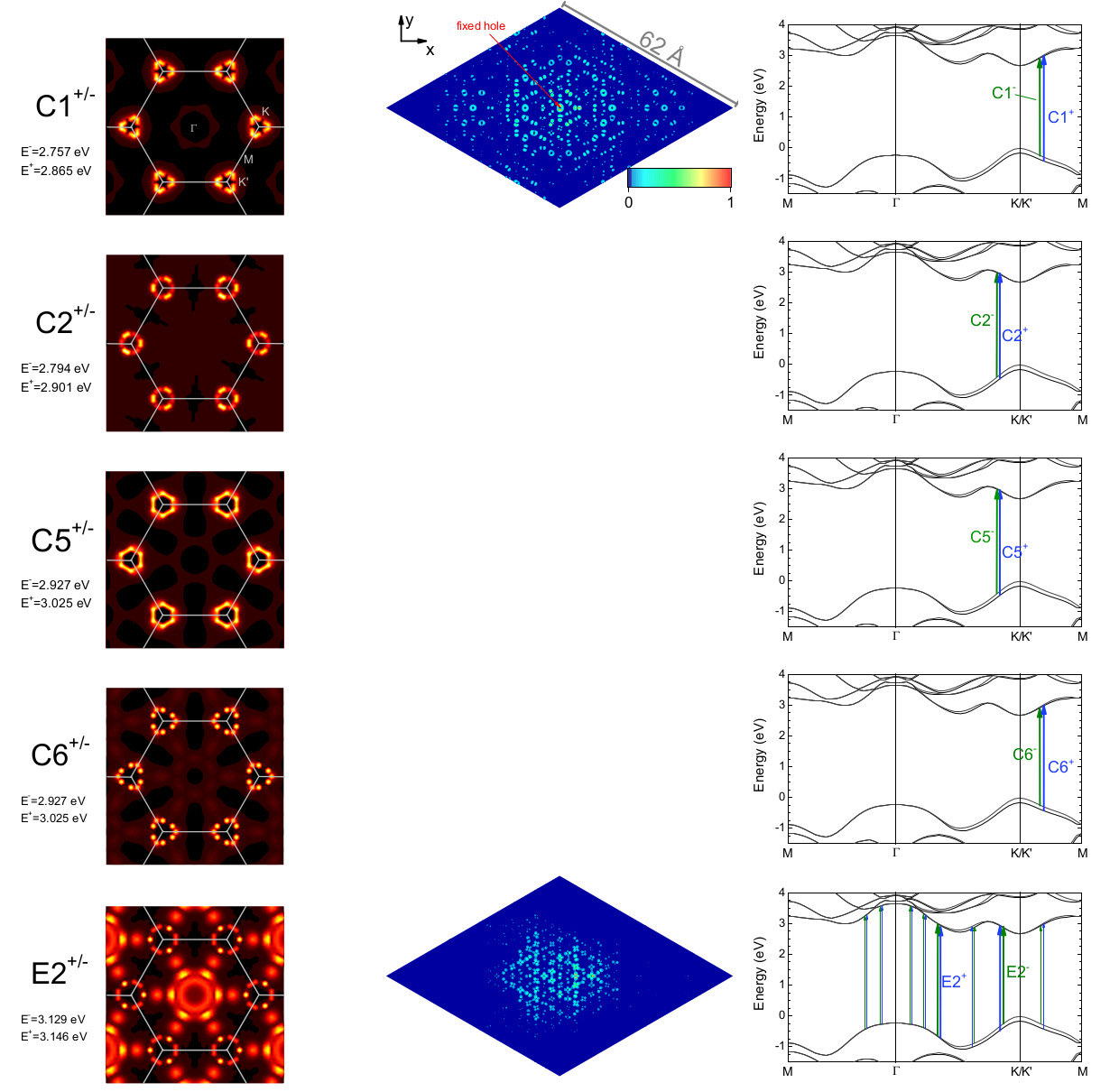}

\caption{\label{fig:MoS2-1L} Reciprocal space representations of selected optical transitions in monolayer MoS$_2$ contributing to the \textsl{C} and {E} peaks in the dielectric fucntion. The real-space wavefunctions of \textsl{C1} and \textsl{E2} are shown on a 20x20 supercell, for the hole being fixed in the center. The positions of the layers in the $x$-$z$ plots are indicated by gray dashed lines. All transitions are sketched in the electronic bandstructures.}
\end{figure}

\newpage
\subsection{1L-MoSe$_2$}
\begin{figure}[hb!]
\centering
\includegraphics*[width=0.92\columnwidth]{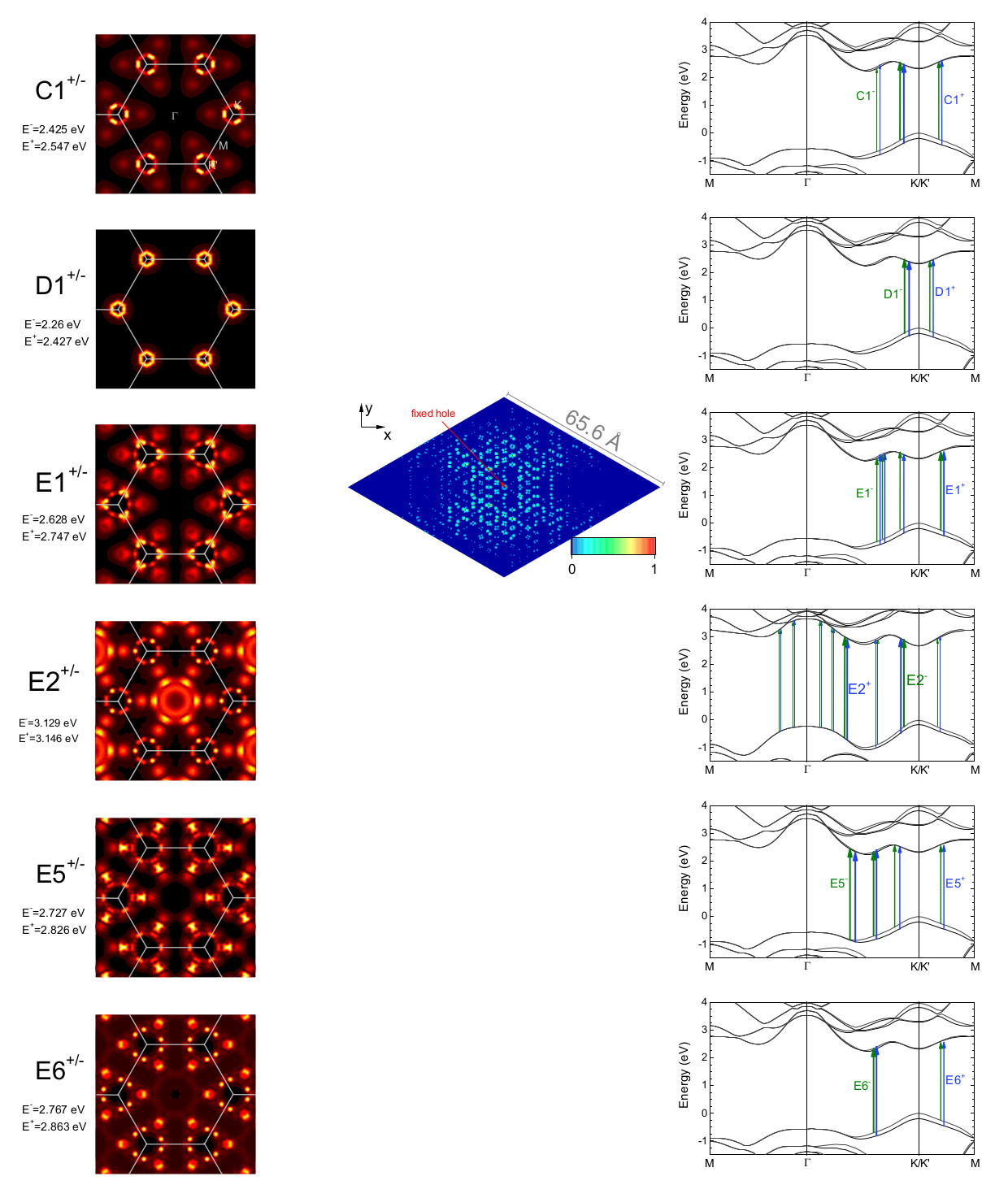}

\caption{\label{fig:MoSe2-1L} Reciprocal space representations of the \textsl{C1}, \textsl{D1} and several \textsl{E} transitions in 1L-MoSe$_2$. As in~\ref{fig:MoS2-1L}, the transitions are sketched in the bandstructures and the real-space wavefunctions are plotted using a 20x20 supercell.}
\end{figure}

Fig.~\ref{fig:MoSe2-1L} contains reciprocal space representations for the \textsl{C1} and \textsl{D1} transitions, which contribute to the \textsl{D} and \textsl{Ca} peaks in the dielectric function, and of a number of transitions that make up the broad \textsl{E} peak.

\newpage
\subsection{1L-MoTe$_2$}
\begin{figure}[hb!]
\centering
\includegraphics*[width=0.93\columnwidth]{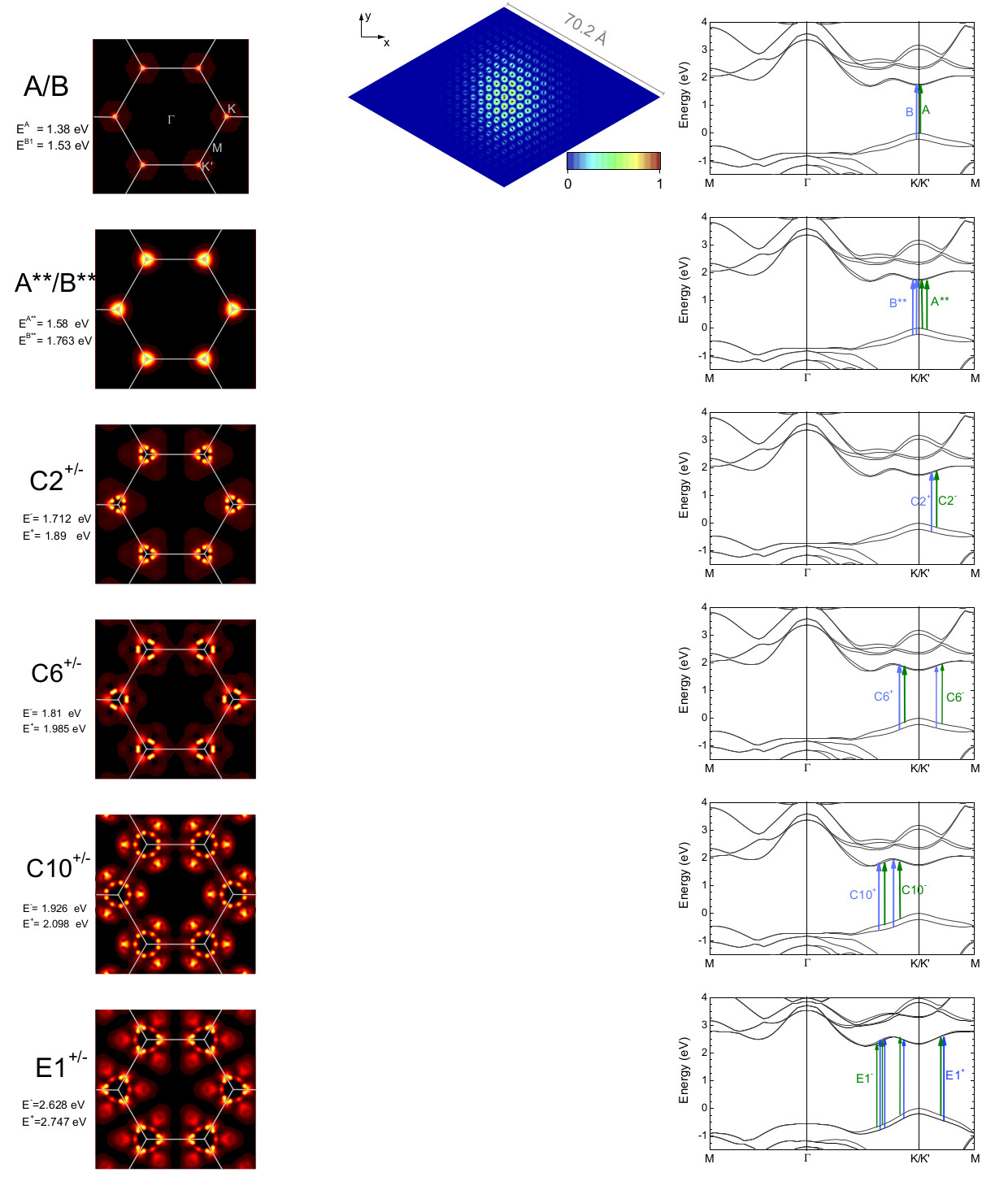}
\caption{\label{fig:MoTe2-1L-1} Reciprocal space representations of additional transitions and sketches in the electronic band structures for monolayer MoTe$_2$. The real-space wavefunction for \textsl{A/B} is shown for the hole being fixed at the central Mo atom.}
\end{figure}
\begin{figure}[hb!]
\centering
\includegraphics*[width=0.93\columnwidth]{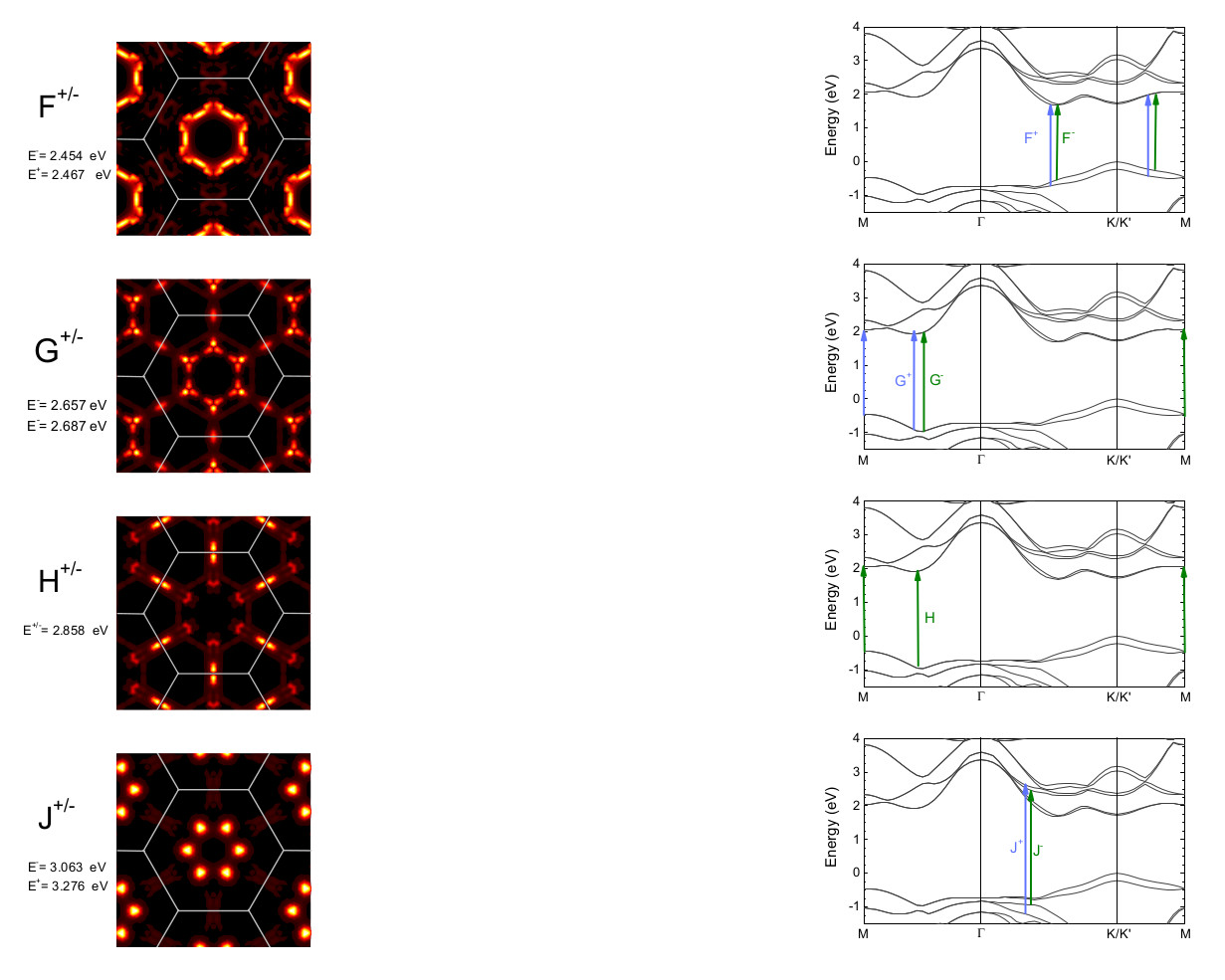}
\caption{\label{fig:MoTe2-1L-2} As Fig.~\ref{fig:MoTe2-1L-1} for the main transitions of the \textsl{F}, \textsl{G}, \textsl{H}, and \textsl{J} peaks in the dielectric function in Fig.~1~(c) of the main text.}
\end{figure}
Fig.~\ref{fig:MoTe2-1L-1} shows the reciprocal space representations of additional transitions (C2, C6 and C10), which contribute to the two split \textsl{Ca} and \textsl{Cb} peaks in the calculated dielectric function. As mentioned in the main text, the \textsl{A/B} and \textsl{A**/B**} transitions correspond to the transitions of the same name in MoS$_2$.
%
%F, G2 and J5: 
In addition to the transitions mentioned in the main text, a number of strong transitions of negligible binding energy show up at higher energies and form the distinct \textsl{F}, \textsl{G}, \textsl{H} and \textsl{J} peaks in the imaginary part of the dielectric function (refer to Fig. 1 (c) of the main text), which might be excited through suitable lasers.

\textsl{F$^{-}$} and \textsl{F$^{+}$} are electronic transitions between $v_1\rightarrow c_1$ and $v_1\rightarrow c_1$, respectively, at the \textit{Q} points and the connection lines, thus forming a hexagonal region in reciprocal space. The \textsl{G} and \textsl{H} transitions, on the other hand, correspond to direct transitions at the \textit{M} point and on the \textit{$\Gamma$}-\textit{M} line. All three features are electronic transitions between the valence and conduction band edges, \emph{i.e.} $v_1\rightarrow c_1$ and $v_1\rightarrow c_1$. Due to the low valence band curvature at the \textit{Q} point, see the bandstructures in Fig.~\ref{fig:MoTe2-1L-2}, we expect the holes to have a significantly larger effective mass compared to that at \textit{K} and \textit{K'}. 

The \textsl{J$^{-}$} peak at 3.06\,eV shows some similarity in shape to the \textsl{C4$^{+/-}$} transition in MoS$_2$ and is localized in reciprocal space at six degenerate points at about $\frac{1}{3}$ along the \textit{$\Gamma$}-\textit{K} line. The high oscillation strength suggests a van-Hove singularity in the joint density of states at these reciprocal space points between $v_3$ and  $c_3$. The corresponding \textsl{J$^{+}$} peak for $v_4\rightarrow$ $c_4$ is split by about 200\,meV.

\newpage
\subsection{2L-MoS$_2$}
Fig.~\ref{fig:MoS2-2L} shows plots for the \textsl{D1} and \textsl{E11} transitions mentioned in the main text. \textsl{D1} corresponds to a direct transition at the $\Gamma$-point, which becomes energetically accessible due to the reduced quantum confinement in the bilayer material compared to monolayer MoS$_2$. As mentioned in the main text, the corresponding wavefunction is very delocalized over real-space, both within the excited MoS$_2$ layer and into the neighbouring layer. The plot shows one of two degenerate transitions: One originating from the upper layer, while the other originates from the lower layer. The reciprocal space representation of \textsl{E11} resembles \textsl{C4} in the monolayer structure, but shows a more complicated real-space representation.
\begin{figure}[hb!]

\centering
\includegraphics*[width=0.92\columnwidth]{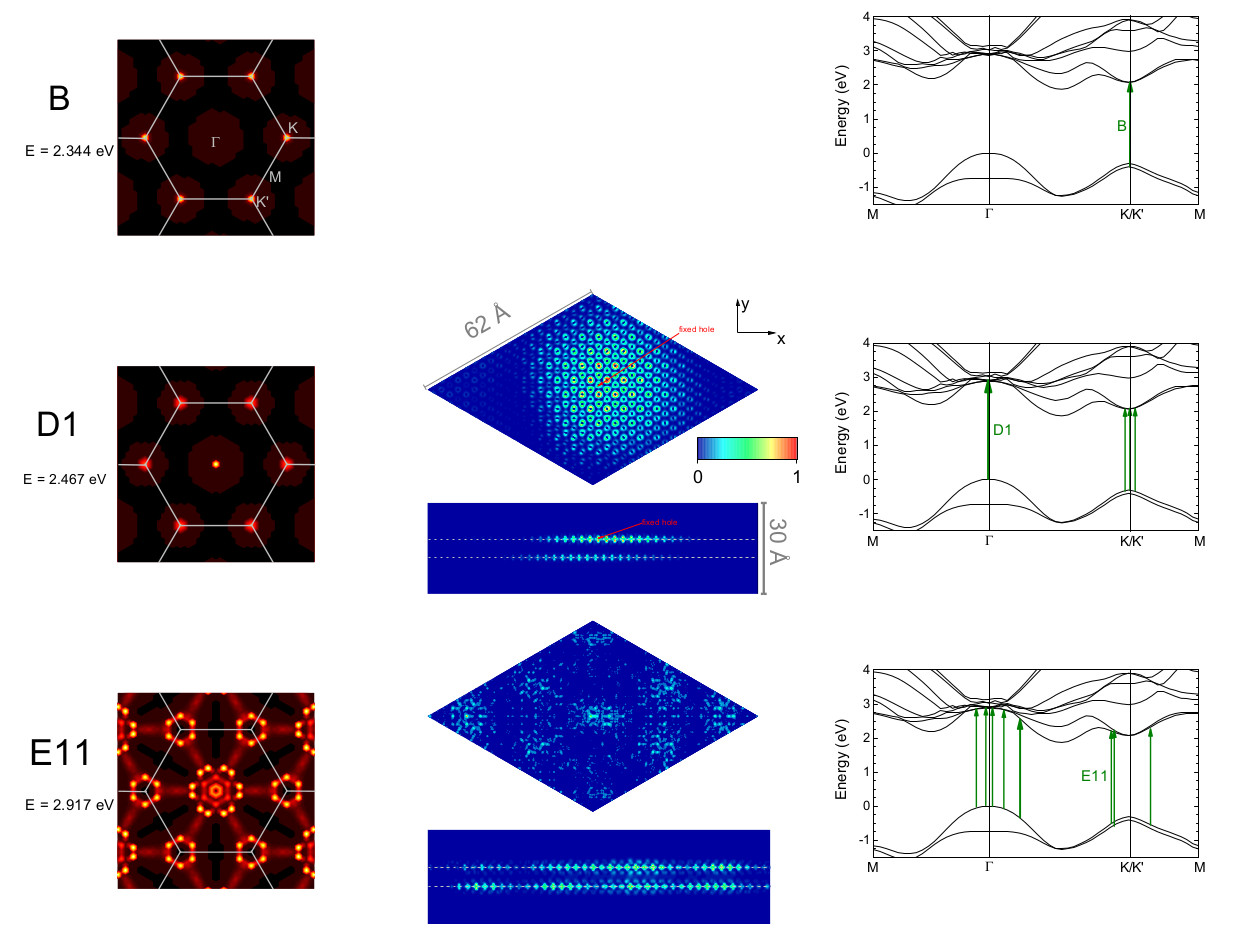}

\caption{\label{fig:MoS2-2L} Reciprocal space representations and real space excitonic wavefunctions (sliced through the upper molybdenum layer, and projected on the $x$-$z$ plane) of selected optical transitions in bilayer MoS$_2$. As for the monolayer materials, we used a 20x20 supercell to expand the excitonic wavefunction in real space and the hole was fixed at the central Mo atom in the upper layer.}
\end{figure}

\newpage

\subsection{2L-MoSe$_2$ and 2L-MoTe$_2$}
\begin{figure}[hb!]
\centering
\includegraphics*[width=0.92\columnwidth]{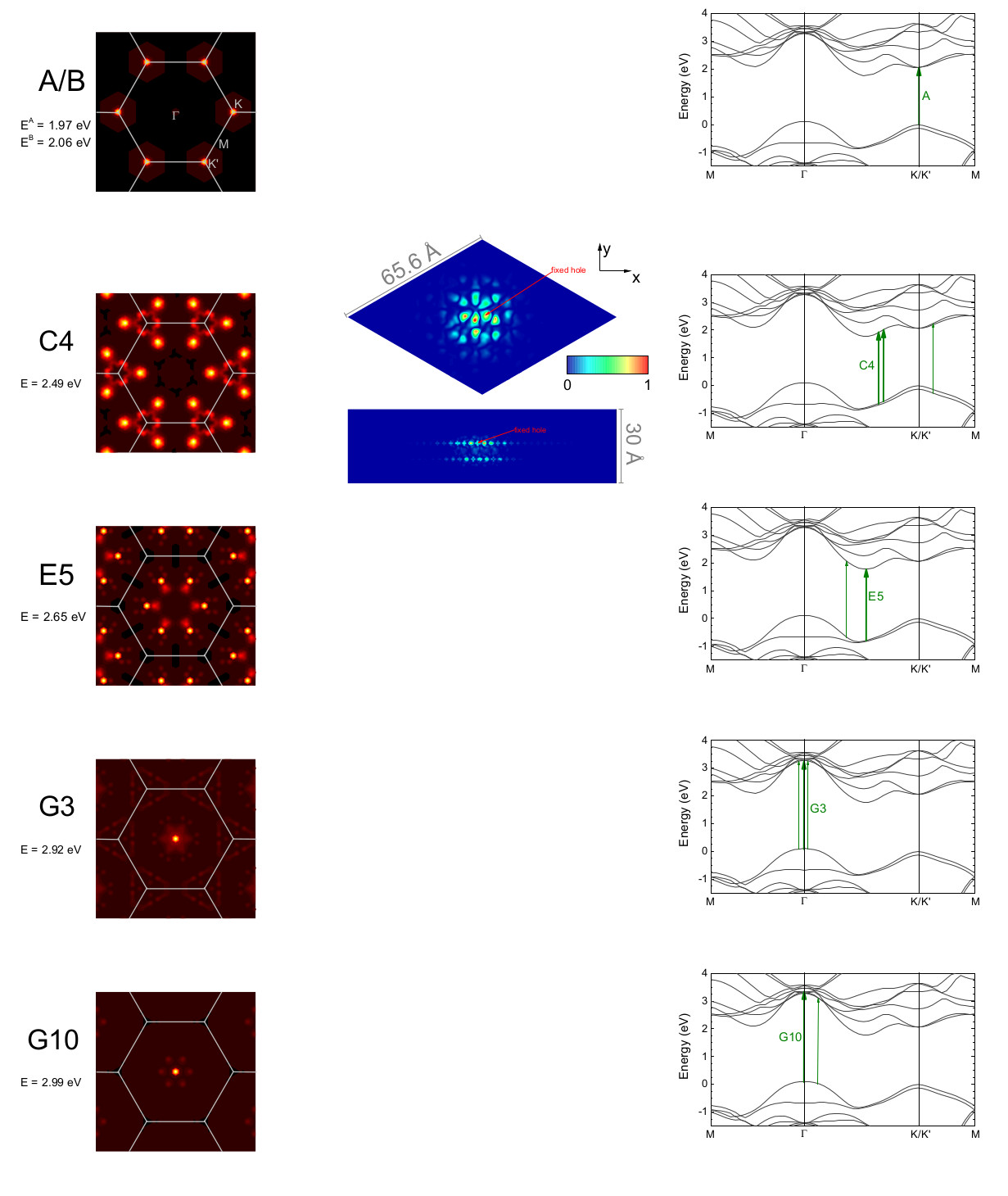}
\caption{\label{fig:MoSe2-2L} Reciprocal space representations and real space excitonic wavefunctions (sliced through the upper molybdenum layer, and projected on the $x$-$z$ plane) of selected optical transitions in bilayer MoSe$_2$.}
\end{figure}
\begin{figure}[hb!]

\centering
\includegraphics*[width=0.92\columnwidth]{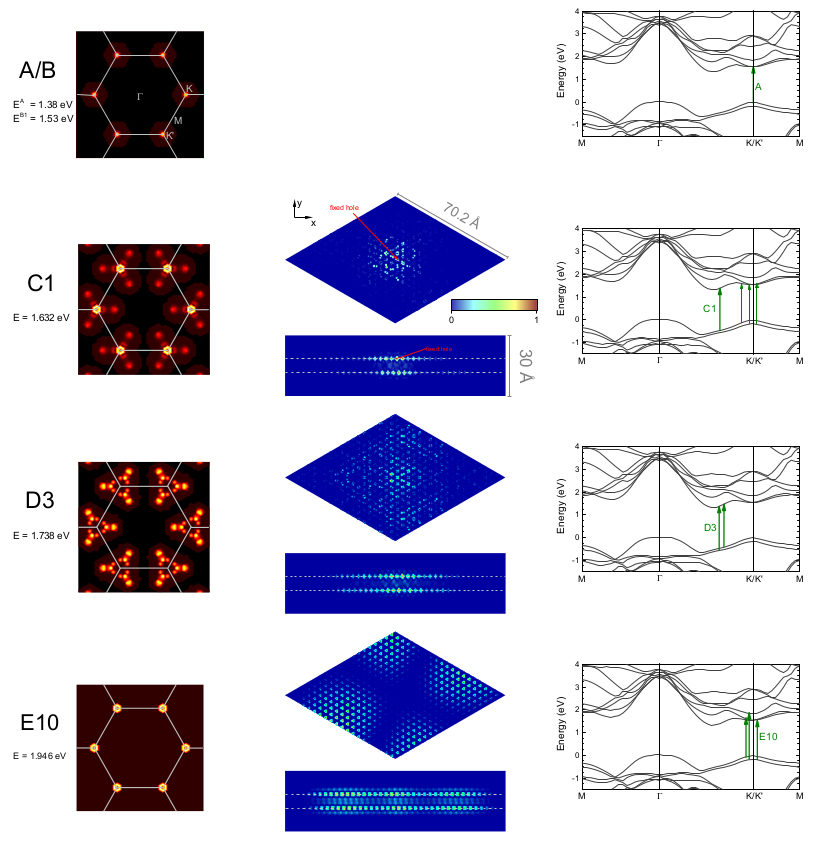}

\caption{\label{fig:MoTe2-2L} Reciprocal space representations and real space excitonic wavefunctions (sliced through the upper molybdenum layer, and projected on the $x$-$z$ plane) of selected optical transitions in bilayer MoTe$_2$.}
\end{figure}

Here, we show selected transitions from our calculations of bilayer MoSe$_2$ (Fig.~\ref{fig:MoSe2-2L}) and MoTe$_2$ (Fig.~\ref{fig:MoTe2-2L}) that did not find space in the main text. The transitions are labeled after Fig.~1~(e) and~(f). In both materials, we find \textsl{A} and \textsl{B} transitions that correspond to the transitions of same name in 2L-MoS$_2$ and are localized in-plane. The \textsl{C4} (MoSe$_2$) and \textsl{C1} (MoTe$_2$) show a great resemblance to both \textsl{C1} in 2L-MoS$_2$ and the dominant \textsl{C} transitions in the monolayer structures: Both transitions have significant off-\emph{K} components that appear at points of high band nesting (see Fig.~\ref{fig:nesting}~(e) and~(f)). 
The excitonic wavefunctions show a distinct localization in real-space and spill over into the non-excited layer, albeit seemingly to a somewhat lesser extent than in 2L-MoS$_2$, compare \textsl{C4} in Fig.~\ref{fig:MoSe2-2L} and \textsl{C1} in Fig.~\ref{fig:MoTe2-2L} to \textsl{C1} in Fig.~7~(c) of the main text. Similar to bilayer MoS$_2$, we find transitions at the $\Gamma$-point for 2L-MoSe$_2$, which form the dominant contributions to the \textsl{G} peak in the dielectric function.

\clearpage
\section{Band nesting}
Here, we show additional plots of the energy gradient differences between first valence and conduction bands for bilayer MoSe$_2$ and MoTe$_2$ and trilayer MoS$_2$ that were derived from the quasiparticle bandstructures. For comparison, the plots for the monolayer systems and bilayer MoS$_2$ were included.

\begin{figure}[h!]
\centering

\includegraphics*[width=\textwidth]{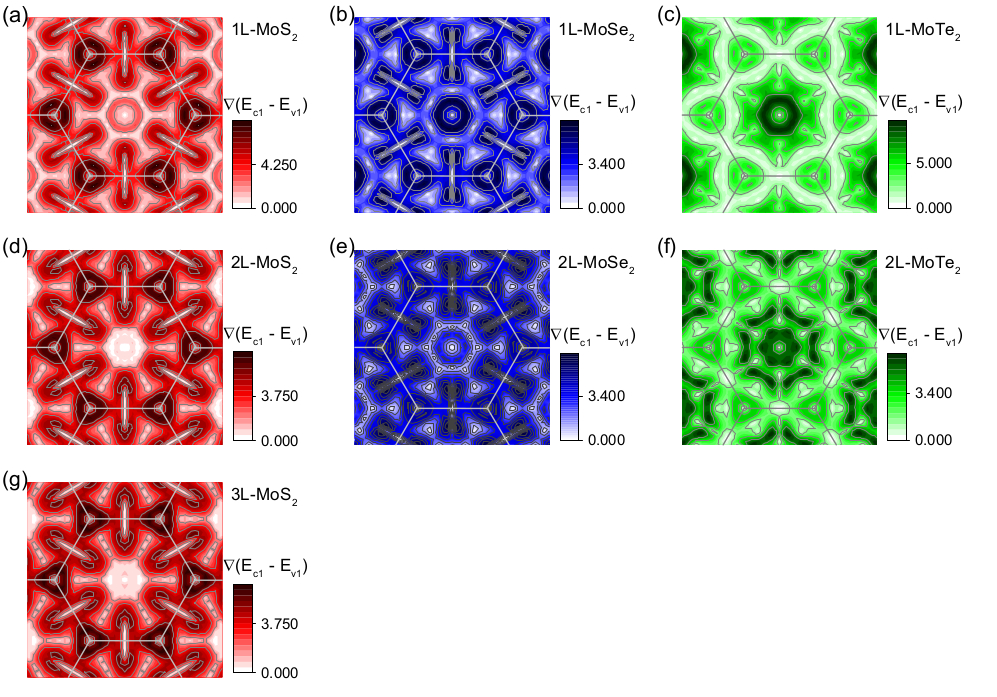}

\caption{\label{fig:nesting} Plots of the difference of the energy gradients between the first valence band, v$_1$, and the first conduction band, c$_1$ for mono- and bilayer MoS$_2$, MoSe$_2$ and MoTe$_2$ and trilayer MoS$_2$. Band nesting occurs in regions with low gradient difference (bright areas). }
\end{figure}
%

%

%%%%%%%%%% end supplementary materials %%%%%%%%%%%%%%%%%%%

\end{document}